\documentclass[useAMS,usenatbib]{mn2e}

\usepackage{times}
\usepackage{graphicx}
\usepackage{subfigure}

\newcommand{\gsim}{\mathrel{\hbox{\rlap{\lower.55ex \hbox {$\sim$}}
                   \kern-.3em \raise.4ex \hbox{$>$}}}}
\newcommand{\lsim}{\mathrel{\hbox{\rlap{\lower.55ex \hbox {$\sim$}}
                   \kern-.3em \raise.4ex \hbox{$<$}}}}

\title[The formation and evolution of pre-stellar discs]{Collapse of a molecular cloud core to stellar densities:  the formation and evolution of pre-stellar discs}
\author[M.R. Bate]{Matthew R. Bate\thanks{E-mail:
mbate@astro.ex.ac.uk}\\ School of Physics, University of Exeter, Stocker
Road, Exeter EX4 4QL \\ Monash Centre for Astrophysics, School of Mathematical Sciences, Monash University, Clayton, Vic 3168, Australia}

\date{\today}
\begin{document}
\maketitle
\begin{abstract}
We report results from radiation hydrodynamical simulations of the collapse of molecular cloud cores to form protostars.  The calculations follow the formation and evolution of the first hydrostatic core/disc, the collapse to form a stellar core, and effect of stellar core formation on the surrounding disc and envelope.  Past barotropic calculations have shown that rapidly-rotating first cores evolve into `pre-stellar discs' with radii up to $\sim 100$~AU that may last thousands of years before a stellar core forms.  We investigate how the inclusion of a realistic equation of state and radiative transfer alters this behaviour, finding that the qualitative behaviour is similar, but that the pre-stellar discs may last $1.5-3$ times longer in the more realistic calculations.  
The masses, radii, and lifetimes of the discs increase for initial molecular cloud cores with faster rotation rates.  In the most extreme case we model, a pre-stellar disc with a mass of 0.22~M$_\odot$ and a radius of $\approx 100$~AU can form in a 1-M$_\odot$ cloud and last several thousand years before a stellar core is formed.  Such large, massive objects may be imaged using ALMA.  Fragmentation of these massive discs may also provide an effective route to binary and multiple star formation, before radiative feedback from accretion onto the stellar core can inhibit fragmentation.  Once collapse to form a stellar core occurs within the pre-stellar disc, the radiation hydrodynamical simulations produce qualitatively different behaviour from the barotropic calculations due to the accretion energy released.  This drives a shock wave through the circumstellar disc and launches a bipolar outflow even in the absence of magnetic fields.
\end{abstract}

\section{Introduction}
\label{introduction}

The first numerical calculations of the collapse of a molecular cloud core to 
stellar core formation, and beyond, were performed by \citet{Larson1969} 
more than four decades ago.  These one-dimensional radiation hydrodynamical 
calculations revealed the main stages of protostar formation:
an almost isothermal collapse until the inner regions become
optically thick, the almost adiabatic formation of the first 
hydrostatic core (typical radius $\approx 5$~AU and initial mass
$\approx 5$~M$_{\rm J}$), the growth of this core as it accreted
from the infalling envelope, the second collapse within this
core triggered by the dissociation of molecular hydrogen,
the formation of the stellar core (initial radius $\approx 2$~R$_\odot$
and mass $\approx 1.5$~M$_{\rm J}$), and, lastly, the long
accretion phase of the stellar core to its final mass.  More recent
one-dimensional studies \citep[e.g.][]{MasInu2000,Commerconetal2011b} have not altered
this picture.

However, in one-dimension, the effects of rotation and magnetic fields
can not be examined.  \cite{Larson1972} began performing two-dimensional
calculation of rotating clouds soon after his ground-breaking one-dimensional calculations.
But multi-dimensional calculations are considerably more time consuming
and it was not until almost two decades after the first one-dimensional models
that the first two dimensional calculations were able to follow the collapse
to the formation of the stellar core \citep{Tscharnuter1987}.  These calculations
showed that with rotation, both the first hydrostatic core and the later-formed 
stellar core became rotationally-flattened and allowed the shock structures 
in the accretion flows to be studied in detail \citep[e.g.][]{Tscharnuteretal2009}.

It was not until nearly three decades after \citeauthor{Larson1969}'s original
calculations that the first three-dimensional calculations following the collapse to
stellar core formation were possible.
\cite{Bate1998} performed three-dimensional calculations of rotating 
molecular cloud cores using a barotropic equation of state 
to mimic the effects of a realistic equation of state and radiative transfer.
He found that if the original molecular cloud core was rotating rapidly
enough, the rotationally-flattened first hydrostatic core could be dynamically
unstable to the growth of non-axisymmetric perturbations.  It deformed into
a bar, followed by the wrapping up of the ends of the bar into a disc with
strong trailing spiral arms.  Gravitational torques removed 
angular momentum and rotational support from the 
inner regions of the first core, quickening the onset
of the second collapse and inhibiting fragmentation
during the second collapse phase.  The development of a bar-mode and spiral
structure is expected for rapidly-rotating
polytropic-like structures \citep[e.g.][]{Durisenetal1986}.  Such instabilities
occur when the ratio of the rotational energy to the magnitude of the
gravitational potential energy of the first core exceeds $\beta=0.274$.
\citeauthor{Bate1998} was also the first to point out that because a
rapidly-rotating first core develops into a disc before the stellar core
forms, the {\it disc forms before the star}.  Rather than hydrostatic cores,
such structures are better described as `pre-stellar discs'.  Once the 
second collapse occurred and a stellar core formed, \citeauthor{Bate1998} 
found an inner disc formed around the stellar core
and grew in radius as material with larger angular momentum fell in.
Bate speculated that if the calculation were able to be followed further,
the outer radius of the inner disc would eventually grow to join on to the
outer disc, leaving the stellar core surrounded by a single large disc.

The next major advance in this field was to begin including the effects of
magnetic fields.  \citet{Tomisaka2002} was the first to follow the collapse
of a molecular cloud core to stellar densities including magnetic fields.
Using two-dimensional calculations, 
he found that both the first core and stellar cores launched magnetically-driven
outflows.  This work has been followed by a number of papers using three-dimensional calculations
to investigate
magnetised collapse and outflows \citep{Machidaetal2005, MacInuMat2006, MacInuMat2008}
and the rotation rates and magnetic field strengths in the first and stellar cores
\citep{MacInuMat2007}.  At the same time, the development of non-axisymmetric
structure at the first core stage and the formation of a pre-stellar disc discovered 
by \cite{Bate1998} and fragmentation has been studied further 
both without \citep{SaiTom2006,SaiTomMat2008, SaiTom2011} and with
magnetic fields \citep{Machidaetal2005,MacInuMat2010,MacMat2011}.
With the increasing capability of observational instruments
and, in particular, looking forward to the capabilities of the Atacama Large Millimeter/submillimeter Array (ALMA),
some of these works also began to investigate the observational signatures of the
first core phase such as their luminosities \citep{SaiTom2006}, spectral energy distributions (SEDs),
and images \citep{SaiTom2011}.

However, all three-dimensional studied listed above
have used barotropic equations of state rather than realistic equations of state with 
radiative transfer.  The first three-dimensional calculations including radiative transfer that followed 
collapse to the point of stellar core formation (but not beyond)  were \citet{WhiBat2006}, using the 
flux-limited diffusion approximation, and \cite{Stamatellosetal2007},
using a radiative cooling approximation.
\cite{Bate2010} investigated the evolution of both first and stellar cores using
radiation hydrodynamical calculations.  He found that the evolution up until stellar core
formation was qualitatively similar to that obtained with a barotropic equation of state,
including bar-mode instabilities and stellar core formation.  However, following the formation
of the stellar core, the energy released has a dramatic effect on the surrounding disc and envelope
and launches a temporary outflow even in the absence of a 
magnetic field.  Recent two-dimensional radiation hydrodynamical simulations that follow the collapse to stellar
densities and well beyond as the stellar core accretes from the infalling envelope \citep{SchTsc2011}
show a similar effect, where the formation of the stellar core drives a temporary outflow.

The most complete three-dimensional calculations of the collapse of molecular cloud cores
to date include radiation hydrodynamics and magnetic fields 
(\citealt{Tomidaetal2010a}a; \citealt{Tomidaetal2010b}b;  \citealt{Commerconetal2011b}).  As with the past
magnetised barotropic calculations, the first core is found to launch outflows, but now the thermodynamics
of the gas is more properly treated.  However, these models only study the evolution of the first
core, and do not follow the second collapse and formation of the stellar core.

In this paper, we extend the study of \cite{Bate2010} who presented the first three-dimensional 
radiation hydrodynamical calculations to follow the collapse beyond the formation of the stellar core.
\citeauthor{Bate2010} only presented results for one particular value of the rotation rate of the molecular cloud core.
Here we investigate how the evolution of the first core or pre-stellar disc varies with different rotation rates, and we
compare the results obtained using radiation hydrodynamics with those obtained using a barotropic equation
of state.  We also study how the thermally-launched outflows following stellar core formation depend on the 
rotation rate, and how the discs evolve.  Finally, we perform many of the calculations at several different 
numerical resolutions to test the effects of numerical resolution and examine convergence.

\section{Computational method}
\label{method}

The calculations presented here were performed 
using a three-dimensional smoothed particle
hydrodynamics (SPH) code based on the original 
version of \citeauthor{Benz1990} 
(\citeyear{Benz1990}; \citealt{Benzetal1990}), but substantially
modified as described in \citet{BatBonPri1995},
\citet*{WhiBatMon2005}, \citet{WhiBat2006},
\cite{PriBat2007}, and 
parallelised using both OpenMP and MPI.

Gravitational forces between particles and a particle's 
nearest neighbours are calculated using a binary tree.  
The smoothing lengths of particles are variable in 
time and space, set iteratively such that the smoothing
length of each particle 
$h = 1.2 (m/\rho)^{1/3}$ where $m$ and $\rho$ are the 
SPH particle's mass and density, respectively
\cite[see ][for further details]{PriMon2007}.  The SPH equations are 
integrated using a second-order Runge-Kutta-Fehlberg 
integrator with individual time steps for each particle
\citep{BatBonPri1995}.
To reduce numerical shear viscosity, we use the
\cite{MorMon1997} artificial viscosity
with $\alpha_{\rm_v}$ varying between 0.1 and 1 while $\beta_{\rm v}=2 \alpha_{\rm v}$
\citep[see also][]{PriMon2005}.

\begin{figure*}
\centering \vspace{-0.5cm}
    \includegraphics[width=17.0cm]{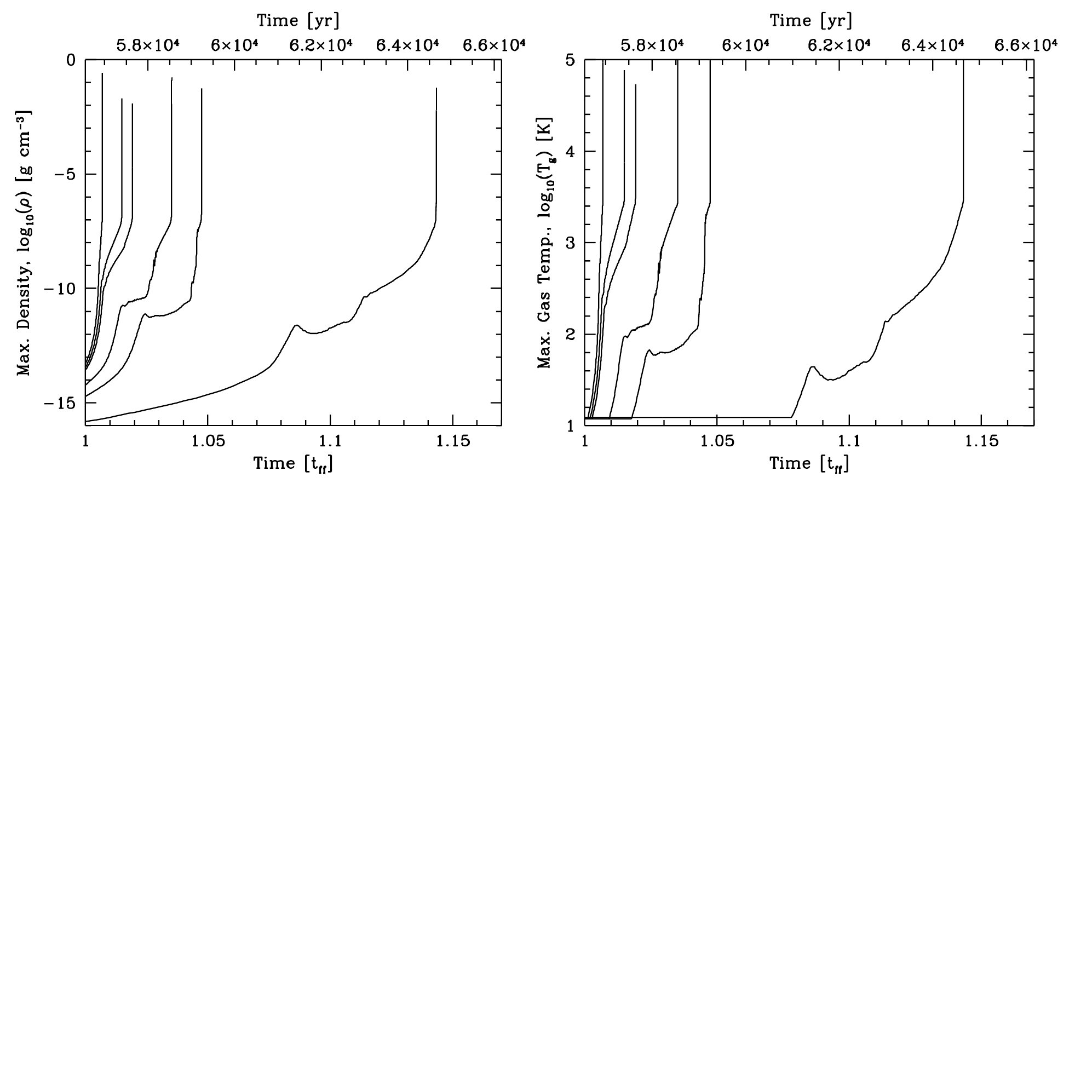}\vspace{-9.5cm}
    \includegraphics[width=17.0cm]{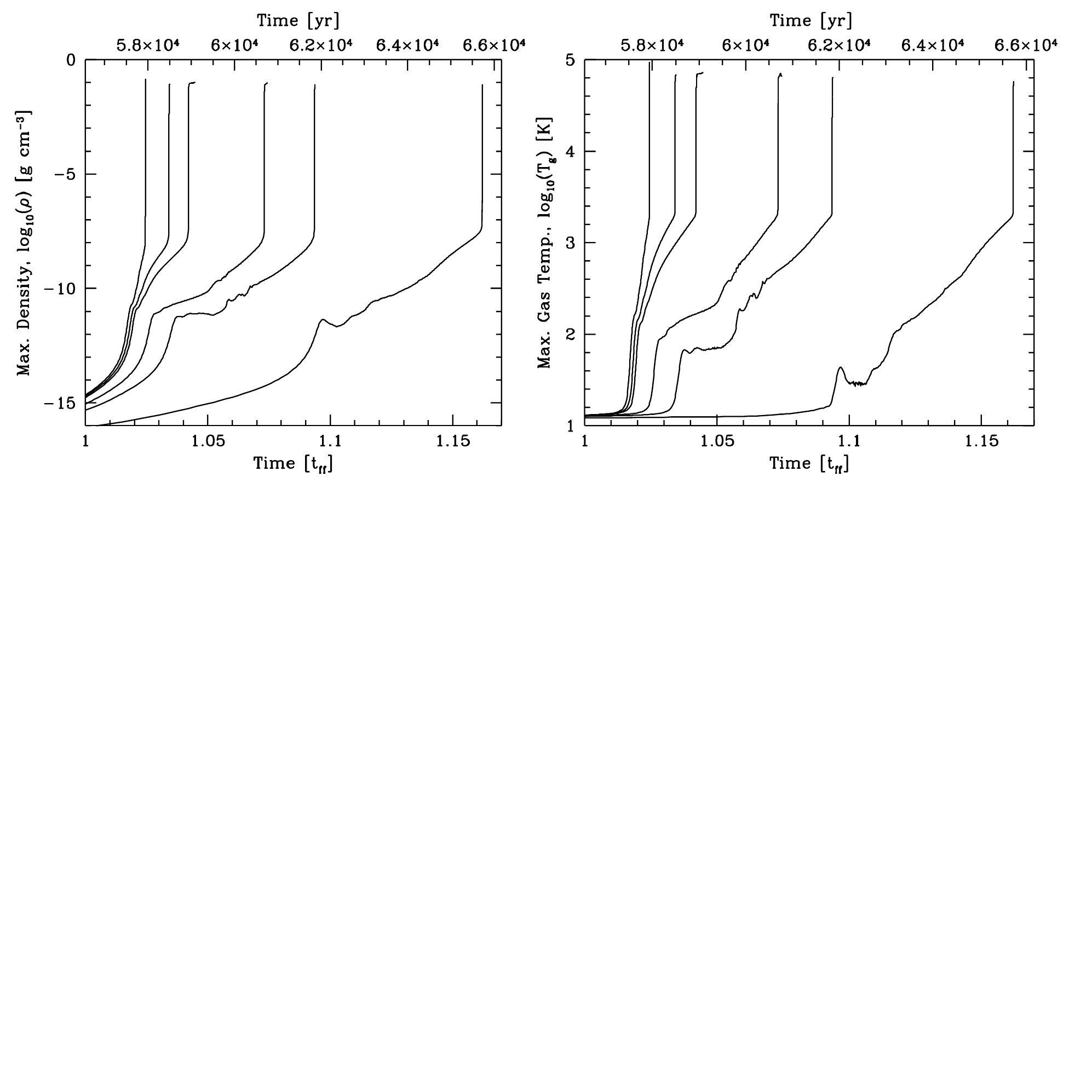}\vspace{-9.5cm}
\caption{The time evolution of the maximum density (left panels) and gas temperature (right panels) during the barotropic calculations (top panels) and the radiation hydrodynamical calculations (bottom panels) of the collapse of molecular cloud cores with different initial rotation rates.  For each panel, the different lines are for cloud cores with $\beta=0,5\times 10^{-4},0.001,0.005,0.01,0.04$ from left to right. The free-fall time of the initial cloud core, $t_{\rm ff}=1.8\times 10^{12}$~s (56,500 yrs).  
Each calculation was performed with $10^6$ SPH particles, except the barotropic calculation with $\beta=0.04$ which used $3\times 10^5$ particles.
}
\label{evolutions}
\end{figure*}

\subsection{Equation of state and radiative transfer}

The calculations presented in this paper, use two different types of equation of state.
The first is a barotropic equation of state, almost identical to that used by \cite{Bate1998},
where the temperature of the gas depends only on its density.
In this case, the thermal pressure of the
gas $p = K \rho^{\eta}$.  
The value of the effective polytropic exponent $\eta$, 
varies with density as
\begin{equation}\label{eta}
\eta = \cases{\begin{array}{rll}
1,	& 			  & \rho \leq 1.0 \times 10^{-13} \cr
7/5, 	& \ 1.0 \times 10^{-13}  < \hspace{-6pt} & \rho \leq 5.7 \times 10^{-8} \cr
1.15, 	& \ 5.7 \times 10^{-8\ } < \hspace{-6pt} & \rho < 1.0 \times 10^{-3} \cr
5/3, 	& 			  & \rho > 1.0 \times 10^{-3}. \cr
\end{array}}
\end{equation}
We take the mean molecular weight of the gas to be $\mu = 2.38$.
The value of $K$ is defined such that when the gas is 
isothermal $K=c_{\rm s}^2$, with the sound speed
$c_{\rm s} = 2.04 \times 10^4$ cm s$^{-1}$,
and the pressure is continuous when the value of $\eta$ changes.

The second type of calculation is performed using radiation hydrodynamics.
In this case, we use an ideal gas equation of state 
$p= \rho T \cal{R}/\mu$, where 
$T$ is the gas temperature, and $\cal{R}$ is the gas constant.  
The thermal evolution takes into account the translational,
rotational, and vibrational degrees of freedom of molecular hydrogen 
(assuming a 3:1 mix of ortho- and para-hydrogen; see
\citealt{Boleyetal2007}).  We also include molecular
hydrogen dissociation, and the ionisations of hydrogen and helium.  
The hydrogen and helium mass fractions are $X=0.70$ and 
$Y=0.28$, respectively.
The contribution of metals to the equation of state and the thermal evolution is neglected.
Two temperature (gas and radiation) radiative transfer in the flux-limited
diffusion approximation is implemented as described by \citet{WhiBatMon2005}
and \citet{WhiBat2006}, except that the standard explicit SPH contributions to the gas energy equation due to the work and artificial viscosity are used when solving the (semi-)implicit energy equations to provide better energy conservation.  We assume solar metallicity gas, using 
the interstellar grain opacity tables of \citet{PolMcKChr1985} and the gas opacity
tables of \citet{Alexander1975} (the IVa King model)  \citep[see][]{WhiBat2006}.

\begin{table}
\begin{tabular}{lccccc}\hline
$\beta$ & Equation  &  \multicolumn{4}{c}{Number of SPH Particles}\\
 & of State & $1\times 10^5$  & $3\times 10^5$ & $1\times 10^6$ & $3\times 10^6$ \\
 \hline
0		& Barotropic 	& & & Y & \\
		& Radiation 	& & & Y & \\
0.0005 	& Barotropic	& & & Y & \\
		& Radiation	& & & Y & \\
0.001	& Barotropic	& & Y & Y & Y \\
		& Radiation	& & Y & Y & Y \\
0.005	& Barotropic	& & Y & Y & \\
		& Radiation	& Y & Y & Y & Y \\
0.01		& Barotropic	& & Y & Y & \\
		& Radiation	& & Y & Y & \\
0.04		& Barotropic	& & Y & & \\
		& Radiation	& & Y & Y & \\ \hline
\end{tabular}
\caption{\label{resolutions} A summary of the type of calculations performed, and their numerical resolutions. From left to right, the columns give the type of initial condition (i.e.\ the values of $\beta$, the ratio of the rotational energy to the magnitude of the gravitational potential energy for the molecular cloud cores), the equations of state used (i.e.\ barotropic equation of state, or radiation hydrodynamics with a realistic equation of state), and which numbers of SPH particles were used to perform calculations.}
\end{table}

\subsection{Initial conditions}
\label{initialconditions}

The initial conditions for the calculations are identical to those 
of \cite{Bate1998} and \cite{Bate2010}.  We follow the collapse 
of initially uniform-density, uniform-rotating, molecular cloud 
cores of mass $M=1 {\rm \ M}_\odot$ and
radius $R=7 \times 10^{16} {\rm \ cm}$.  
The free-fall time of the initial clouds is $t_{\rm ff}=1.785\times 10^{12}$~s (56,500 yrs). 
The ratios of the thermal
and rotational energies to the magnitude of the gravitational 
potential energy are $\alpha = 0.54$ and $\beta$, respectively.  We have performed
calculations with $\beta=0$ (not rotating),
and $\beta = 5\times 10^{-4}, 0.001, 0.005$, 0.01 and 0.04.
Typically, each calculation was performed twice, once using the barotropic equation of 
state, and again using radiation hydrodynamics.

To satisfy the resolution criterion of \cite{BatBur1997} that the 
minimum Jeans mass during the calculation contains at least 
$\approx 2 N_{\rm neigh} = 100$ particles, we require at 
least $1 \times 10^5$ equal-mass particles.  To test for convergence, 
we performed calculations using $1 \times 10^5$, $3 \times 10^5$, $1 \times 10^6$,
and $3\times 10^6$ equal-mass SPH particles.  Only the $\beta=0.005$ radiation hydrodynamical
calculation was performed with all these resolutions, but most calculations
were performed with at least two different resolutions (see Table \ref{resolutions}).  
Calculations using the highest resolution of $3 \times 10^6$ SPH particles were only 
performed for the $\beta=0.001$ cases and radiation hydrodynamical $\beta=0.005$ case.  
In this paper, unless otherwise stated, when a particular set of 
initial conditions and equation of state is discussed or a figure is presented, 
the calculation performed with the highest resolution is used.
We discuss the degree to which the simulations are numerically converged in Appendix A
and at various points in Section \ref{results}.

The calculations were performed on the University of Exeter 
Supercomputer, an SGI Altix ICE 8200.

\begin{figure*}
\centering \vspace{-0.5cm}
    \includegraphics[width=19.0cm]{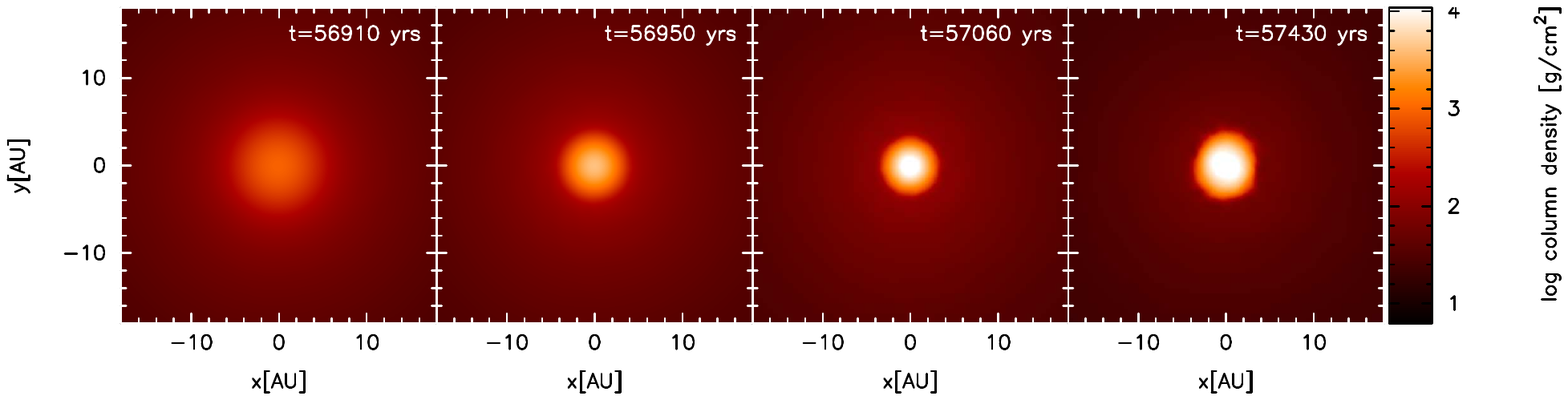}\vspace{-20.7cm}
    \includegraphics[width=19.0cm]{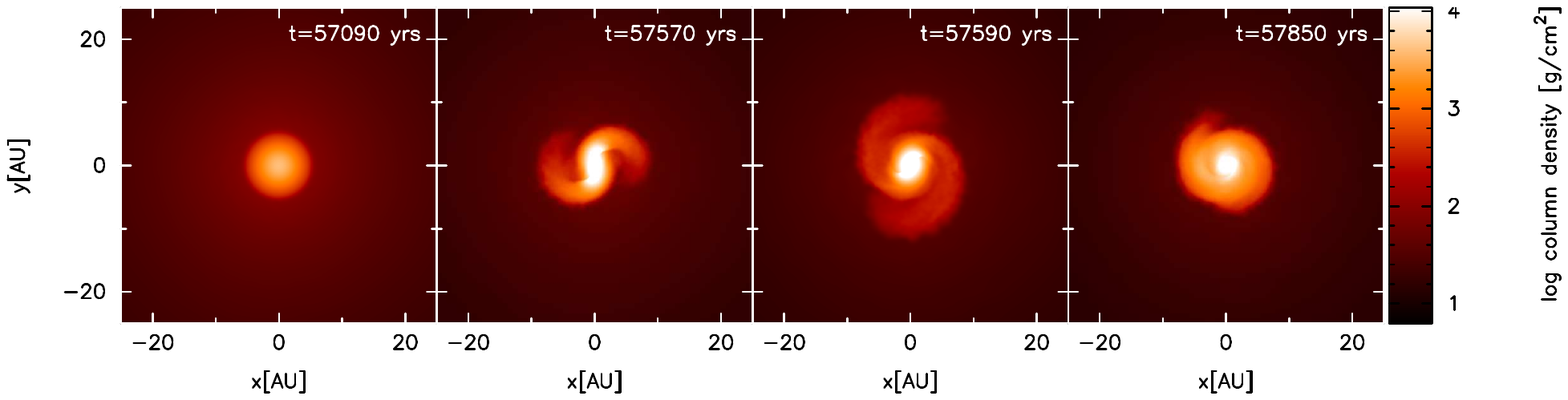}\vspace{-20.7cm}
    \includegraphics[width=19.0cm]{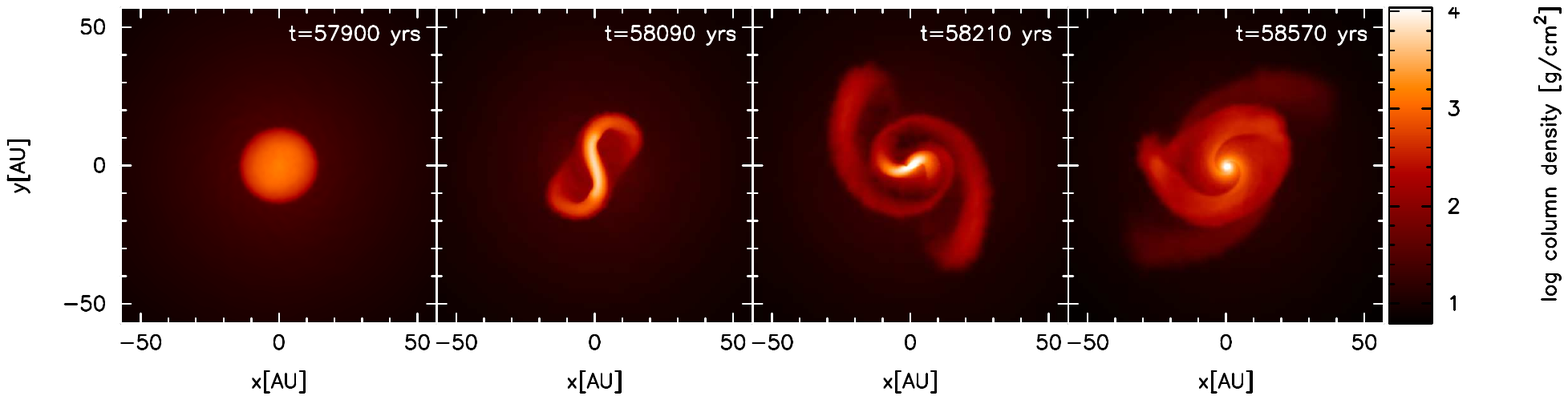}\vspace{-20.7cm}
    \includegraphics[width=19.0cm]{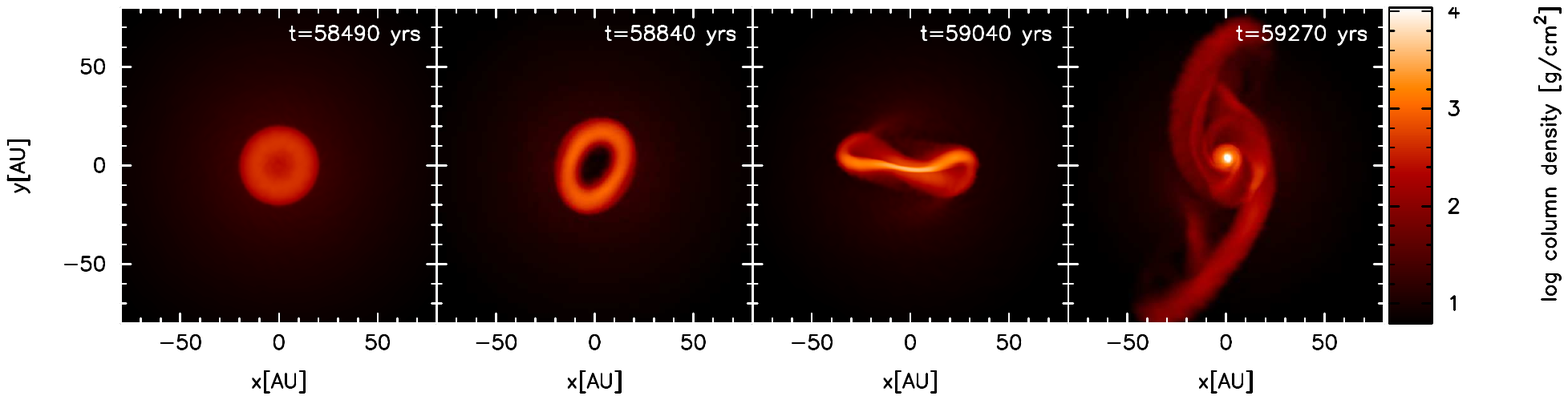}\vspace{-20.7cm}
    \includegraphics[width=19.0cm]{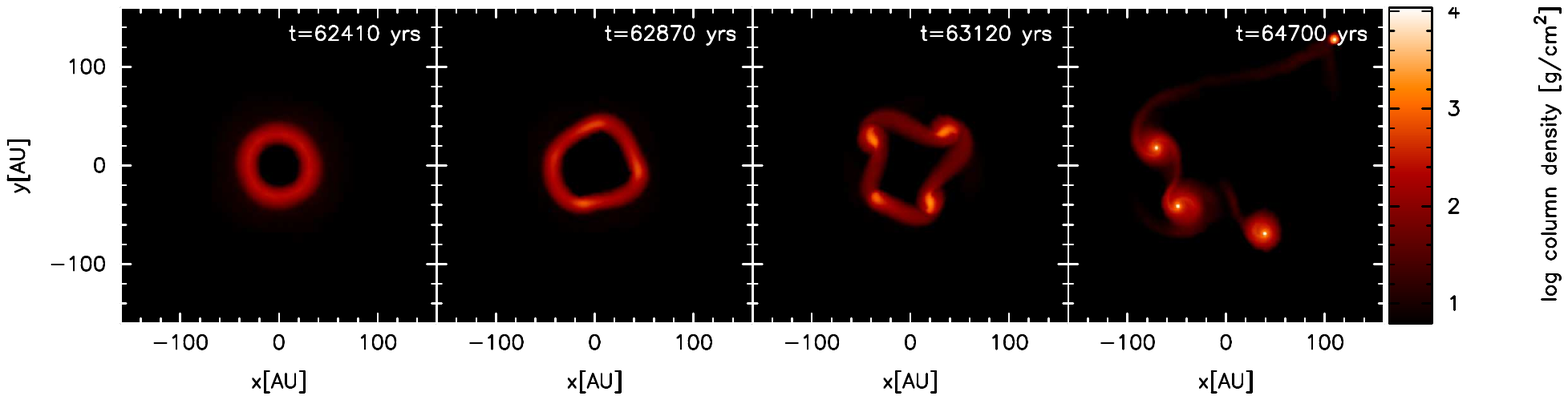}\vspace{-19.5cm}
\caption{Snapshots of the column density viewed parallel to the rotation axis during the evolution of the barotropic calculations of the collapse of molecular cloud cores with different initial rotation rates.  From top to bottom, the different rows are for cloud cores with $\beta=5\times 10^{-4},0.001,0.005,0.01,0.04$. Note that the spatial scale is different for each row, with each panel measuring $1600 \sqrt{\beta}$ AU across (i.e. from 36 to 320 AU).  The free-fall time of the initial cloud core, $t_{\rm ff}=1.8\times 10^{12}$~s (56,500 yrs).  
Each calculation was performed with $10^6$ SPH particles, except for the $\beta=0.001$ case which used $3 \times 10^6$ SPH particles and the $\beta=0.04$ case which used $3 \times 10^5$ particles.  The evolution is almost identical when using $3\times 10^5$ particles or more, but the latter are slightly more detailed.
}
\label{images_D_barotropic}
\end{figure*}

\begin{figure*}
\centering \vspace{-0.5cm}
    \includegraphics[width=19.0cm]{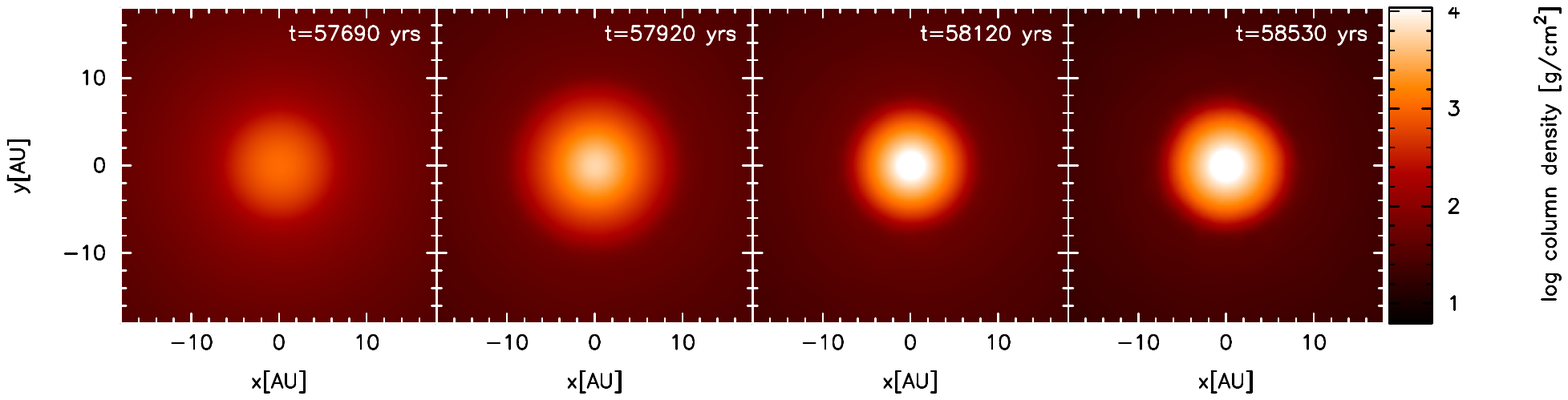}\vspace{-20.7cm}
    \includegraphics[width=19.0cm]{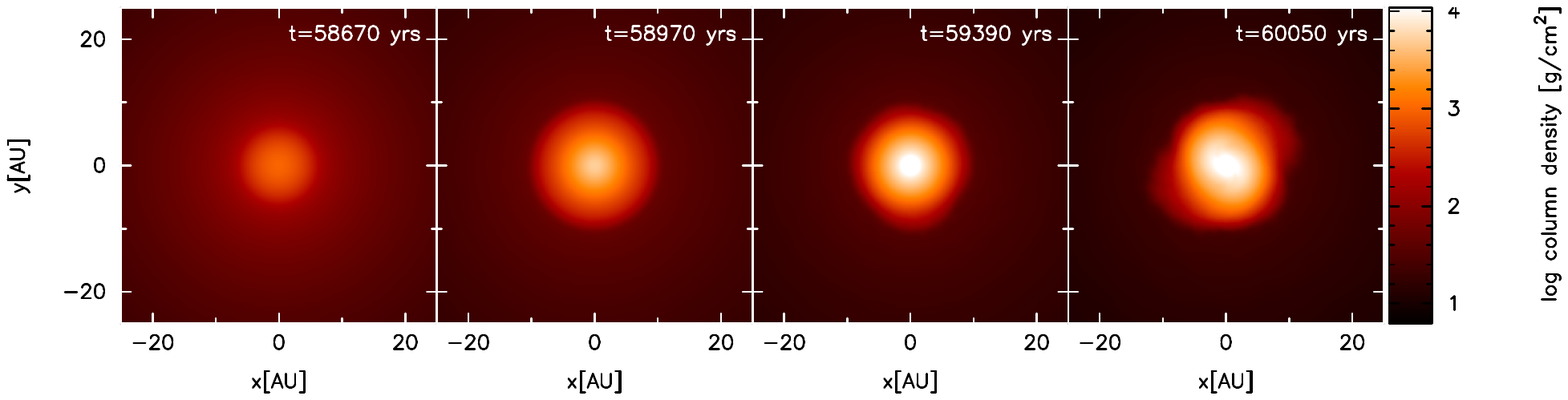}\vspace{-20.7cm}
    \includegraphics[width=19.0cm]{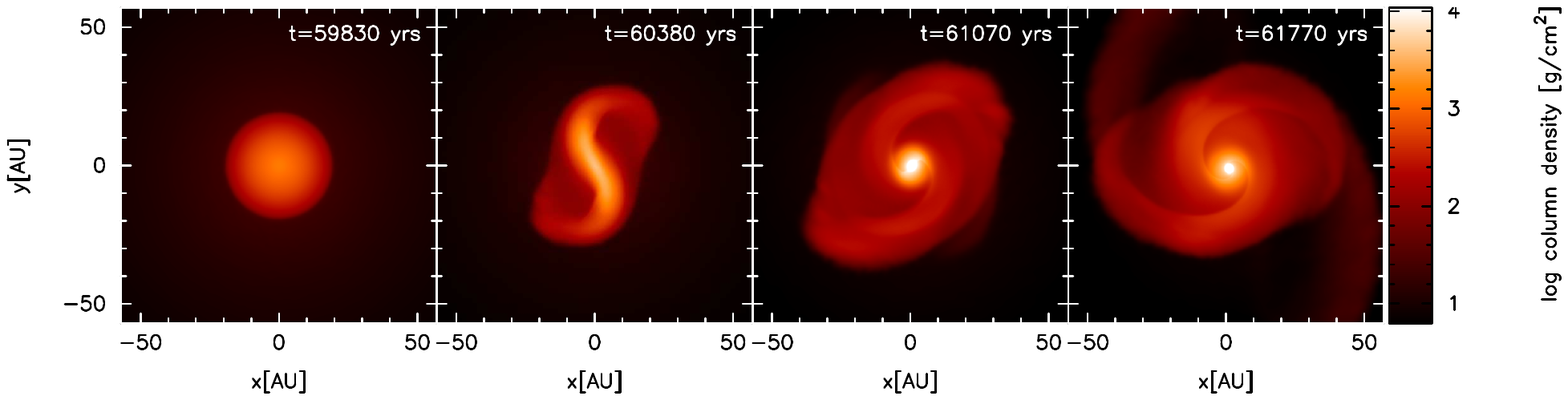}\vspace{-20.7cm}
    \includegraphics[width=19.0cm]{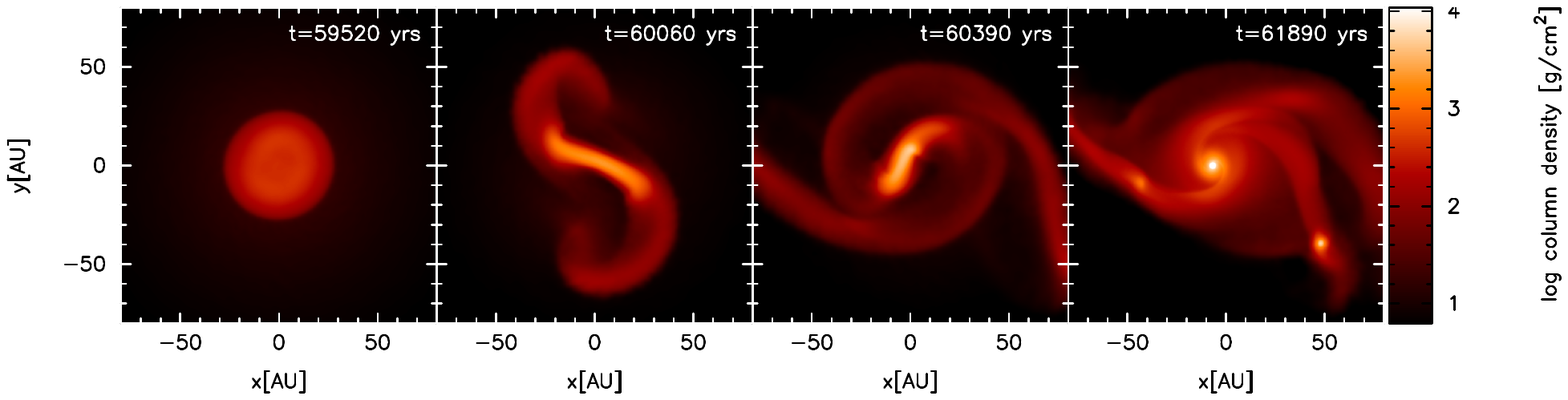}\vspace{-20.7cm}
    \includegraphics[width=19.0cm]{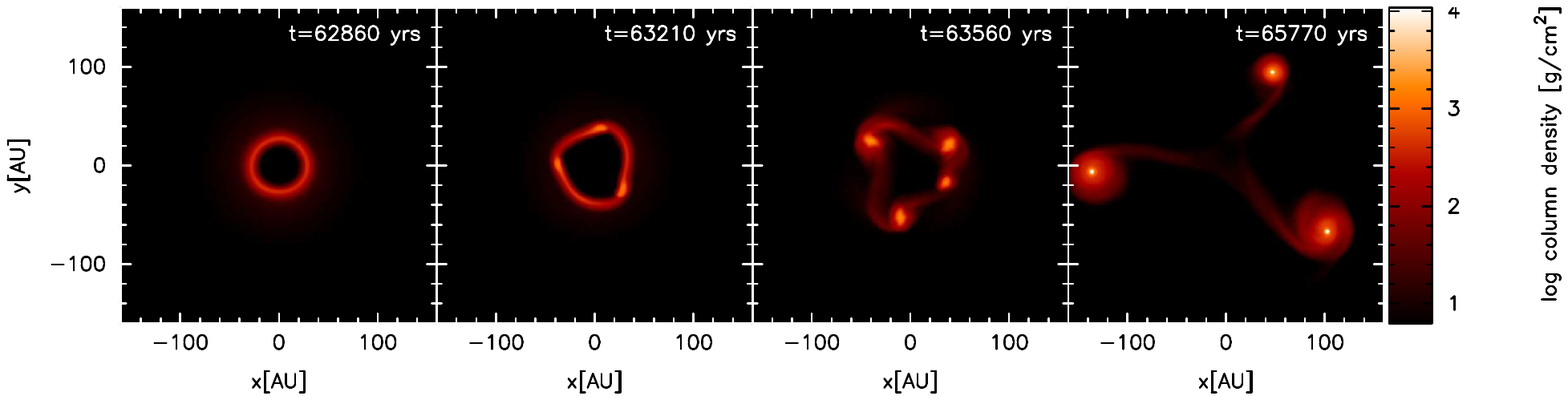}\vspace{-19.5cm}
\caption{Snapshots of the column density viewed parallel to the rotation axis during the evolution of the radiation hydrodynamical calculations of the collapse of molecular cloud cores with different initial rotation rates.  From top to bottom, the different rows are for cloud cores with $\beta=5\times 10^{-4},0.001,0.005,0.01,0.04$. Note that the spatial scale is different for each row, with each panel measuring $1600 \sqrt{\beta}$ AU across (i.e. from 36 to 320 AU).  The free-fall time of the initial cloud core, $t_{\rm ff}=1.8\times 10^{12}$~s (56,500 yrs).  
Each calculation was performed with $10^6$ SPH particles, except for the $\beta=0.001$ and $\beta=0.005$ cases which used $3 \times 10^6$ SPH particles.  The evolution is almost identical when using $10^6$ or $3 \times 10^6$ particles, but the latter are slightly more detailed (see Appendix A).
}
\label{images_xyD}
\end{figure*}

\begin{figure*}
\centering \vspace{-0.5cm}
    \includegraphics[width=19.0cm]{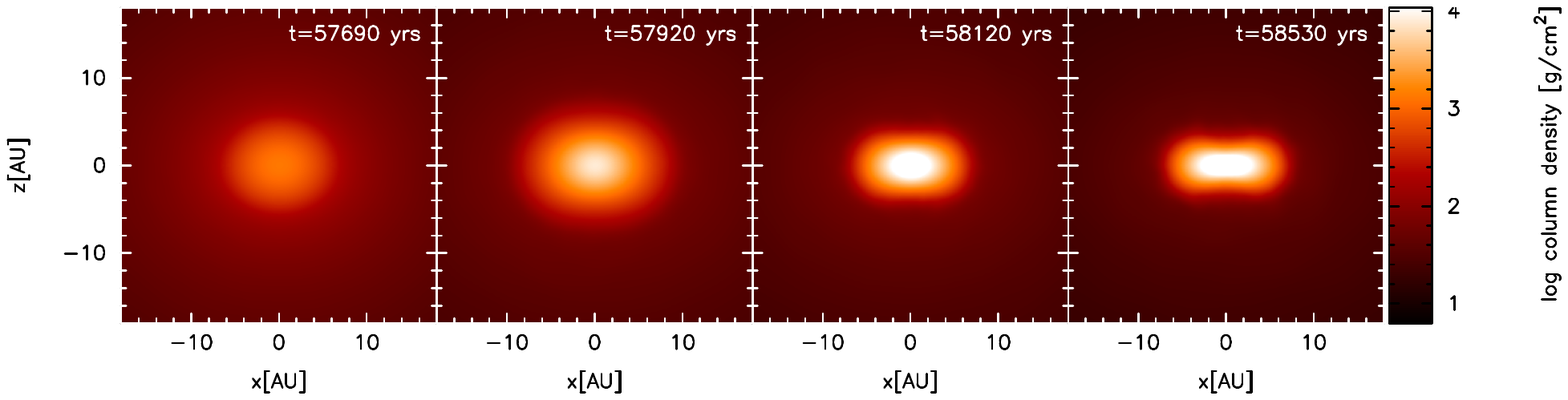}\vspace{-20.7cm}
    \includegraphics[width=19.0cm]{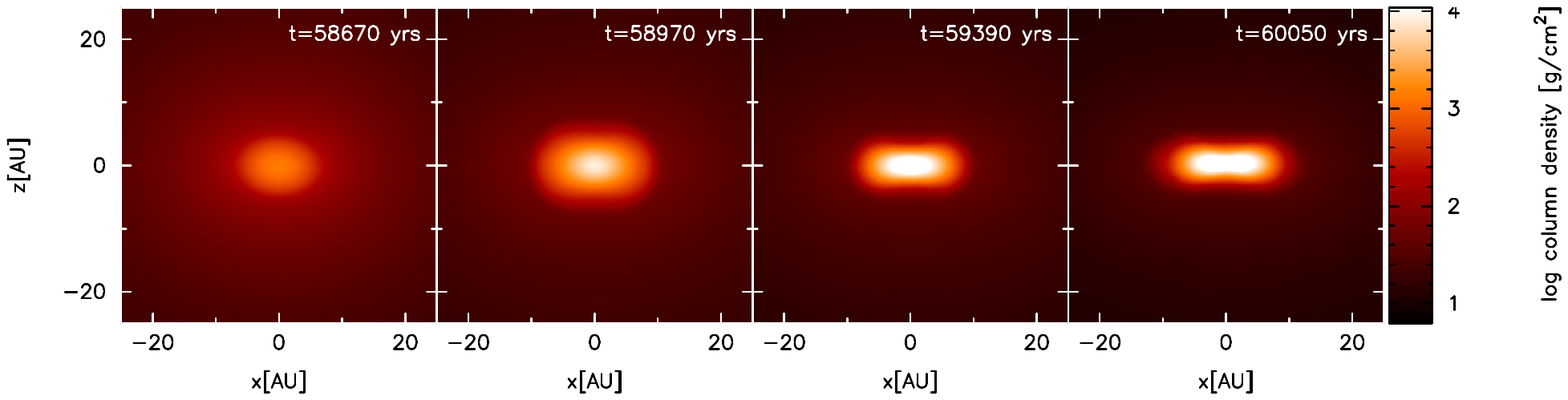}\vspace{-20.7cm}
    \includegraphics[width=19.0cm]{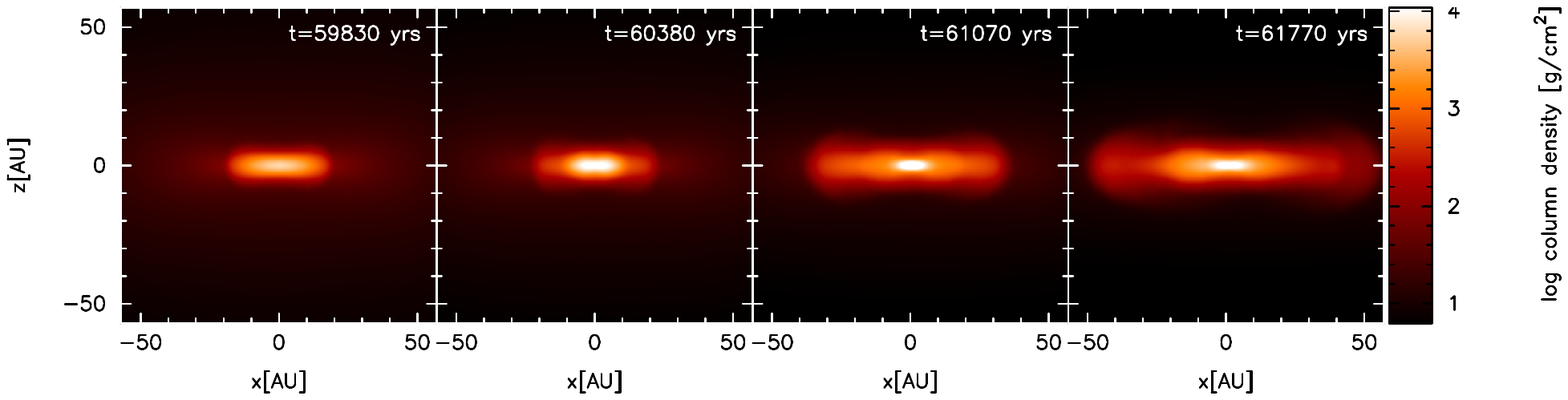}\vspace{-20.7cm}
    \includegraphics[width=19.0cm]{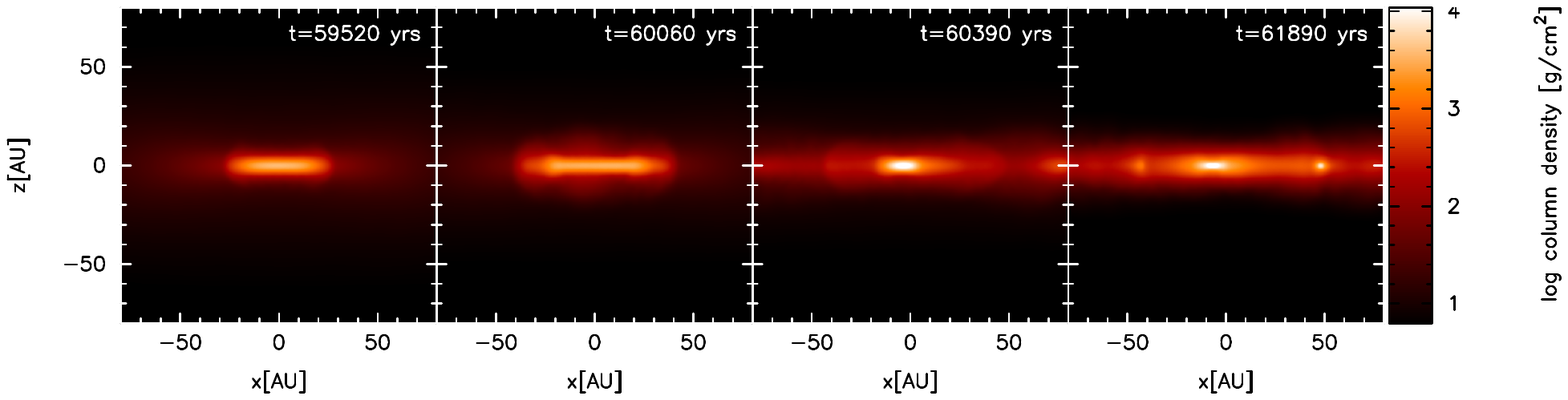}\vspace{-20.7cm}
    \includegraphics[width=19.0cm]{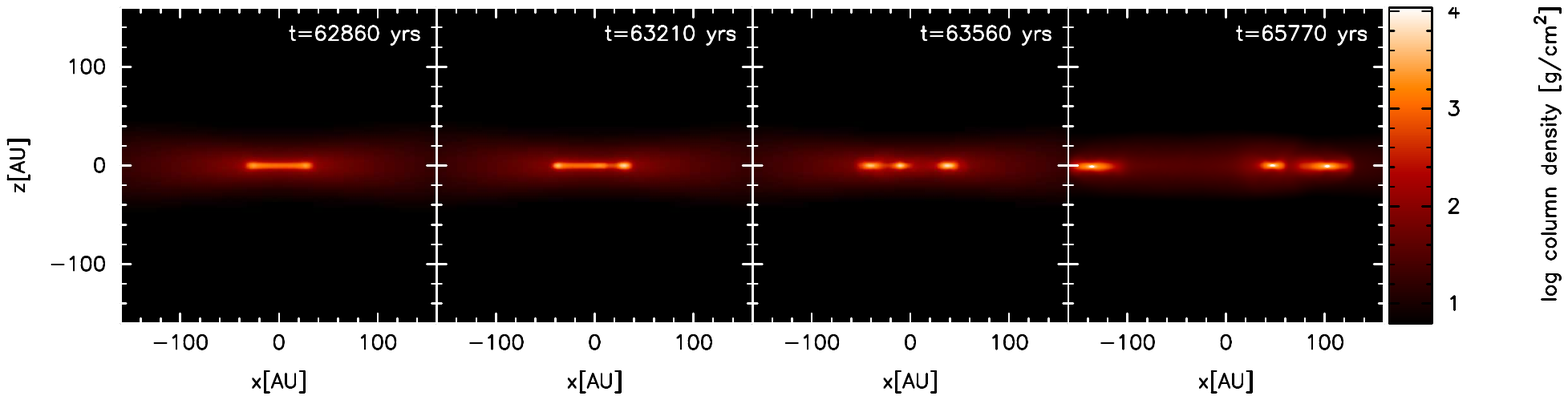}\vspace{-19.5cm}
\caption{Snapshots of the column density viewed perpendicular to the rotation axis during the evolution of the radiation hydrodynamical calculations of the collapse of molecular cloud cores with different initial rotation rates.  From top to bottom, the different rows are for cloud cores with $\beta=5\times 10^{-4},0.001,0.005,0.01,0.04$. Note that the spatial scale is different for each row, with each panel measuring $1600 \sqrt{\beta}$ AU across (i.e. 36, 50, 114, 160, or 320 AU).  The free-fall time of the initial cloud core, $t_{\rm ff}=1.8\times 10^{12}$~s (56,500 yrs).  
Each calculation was performed with $10^6$ SPH particles, except for the $\beta=0.001$ and $\beta=0.005$ cases which used $3 \times 10^6$ SPH particles.  The evolution is almost identical when using $10^6$ or $3 \times 10^6$ particles, but the latter are slightly more detailed (see Appendix A).
}
\label{images_xzD}
\end{figure*}

\begin{figure*}
\centering \vspace{-0.5cm}
    \includegraphics[width=17.5cm]{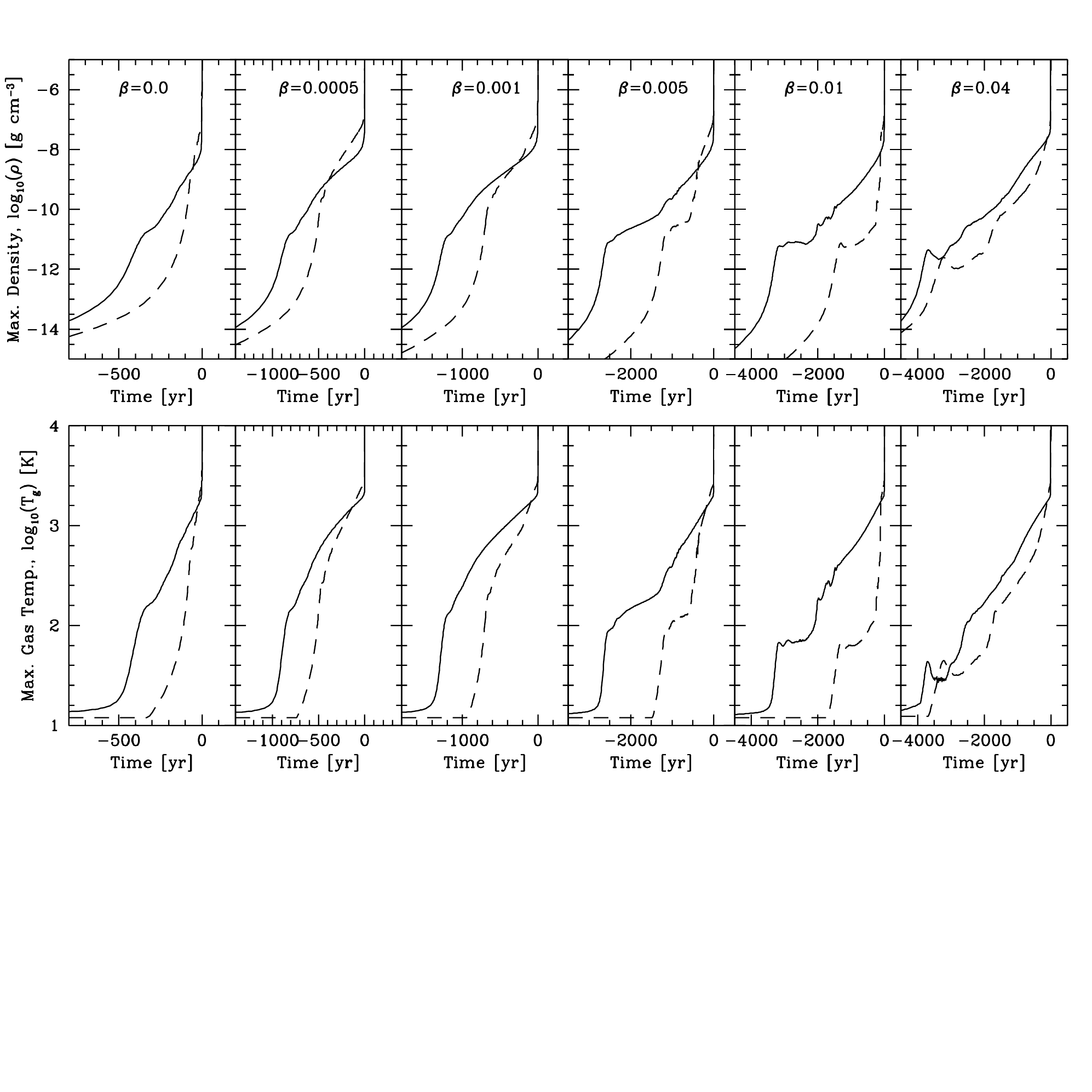}\vspace{-5.3cm}
\caption{The time evolution of the maximum density (upper panels) and gas temperature (lower panels) during the radiation hydrodynamical calculations of the collapse of molecular cloud cores with different initial rotation rates.  From left to right, the different panels are for cloud cores with $\beta=0,5\times 10^{-4},0.001,0.005,0.01,0.04$. The dashed lines give the evolution using the barotropic equation of state, while the solid lines give the evolution using radiation hydrodynamics.  Time is set to zero at the end of the second dynamic collapse phase when the density reaches $10^{-3}~{\rm g}~{\rm cm}^{-3}$ which allows the length of the first hydrostatic core phases to be compared.  It can be seen that the length of the first core phase increases with increasing rotation rate, but also that the use of radiation hydrodynamics and a realistic equation of state lengthens the first core phase relative to calculations that use the barotropic equation of state.  Each calculation was performed with $10^6$ SPH particles, except the barotropic calculation with $\beta=0.04$ which used $3\times 10^5$ particles.}
\label{first_core_time}
\end{figure*}

\section{Results}
\label{results}

The collapse of each molecular cloud core up until the formation of the stellar
core proceeds in a manner that is qualitatively similar to those reported from previous 
three-dimensional calculations without magnetic fields and using barotropic equations of state
\citep{Bate1998, SaiTom2006, SaiTomMat2008, MacInuMat2010, MacMat2011} for both the
barotropic and radiation hydrodynamical calculations presented here.

Fig.~\ref{evolutions}, gives the evolution with time of the maximum density 
and temperature for clouds with different initial rotation rates, with each calculation performed using
$10^6$ particles (except for the baratropic $\beta=0.04$ case).
The initial collapse is isothermal
until the maximum density exceeds $10^{-13}$~g~cm$^{-3}$ using the barotropic 
equation of state.  Using radiation hydrodynamics the evolutions to this density are
also almost isothermal.  However, slight heating when using radiation 
hydrodynamics does slow the collapse marginally, leading to the times taken to
form the first hydrostatic cores being slightly longer ($\approx 0.01-0.015~{\rm t}_{\rm ff}$)
than in the corresponding barotropic calculations.

In the radiation hydrodynamical calculations, as the heating rate in the 
central regions exceeds the rate at which the gas can cool, the gas begins
to heat up and the collapse enters an almost adiabatic phase
where the temperature rises as the gas is compressed.  This is mimicked in the barotropic
calculations by the change in the value of $\eta$ from 1 to $7/5$ (equation \ref{eta}).
This increasing temperature leads to the
formation of a pressure-supported `first hydrostatic core' \citep{Larson1969}, which can be seen in Fig.~\ref{evolutions} when the initial collapse
stalls with central (maximum) densities $\sim 10^{-11}$~g~cm$^{-3}$ and 
temperatures of $\approx 30-120$~K, depending on the degree of
rotational support (i.e. cores that rotate more quickly have lower
maximum temperatures).  Again, apart from the small offset in the time of first core formation,
the barotropic and radiation hydrodynamical calculations give very similar results to this point.

Without rotation ($\beta=0$), the first core has an initial mass of $\approx 5$ Jupiter masses (M$_{\rm J}$)
and a radius of $\approx 5$~AU \citep[in agreement with][]{Larson1969}.  However, with higher initial rotation rates 
of the molecular cloud core,
the first cores become progressively more oblate (Figures \ref{images_D_barotropic}, \ref{images_xyD}, and \ref{images_xzD}).  For example, with $\beta = 0.005$ using radiation hydrodynamics, before the onset of dynamical instability, the first core has a radius of $\approx 20$~AU and a major to minor axis ratio of $\approx$4:1 (first panel in each third row of Figs.~\ref{images_xyD} and \ref{images_xzD}).  
With $\beta=0.01$, the first core has a radius of $\approx 30$~AU and a major to minor axis ratio of $\approx$6:1 (fourth rows of these figures).
Thus, for the higher rotation rates, the first core is actually a pre-stellar disc, without a central
object.  As pointed out by \cite{Bate1998}, \cite{MacInuMat2010}, and \cite{Bate2010}, 
the disc actually forms {\it before} the star.  For the very highest rotation rates ($\beta=0.04$),
the first core actually takes the form of a torus or ring
(first panel in each bottom row of Figs.~\ref{images_xyD} and \ref{images_xzD}) 
in which the central density is lower than the maximum density.

\subsection{Slowly-rotating first hydrostatic cores}
\label{slowly_rotating}

The evolution of the first core up until the point of stellar core formation depends on its rotation rate.
Non-rotating and slowly rotating cores evolve as they accrete mass from the
surrounding infalling envelope with their central densities and temperatures increasing
(Figs.~\ref{evolutions} and \ref {first_core_time}, calculations with $\beta \le 0.001$).  In the radiation hydrodynamical 
calculations, when the
central temperature exceeds $\approx 2000$~K, molecular 
hydrogen begins to dissociate, leading to a second 
hydrodynamic collapse deep within the first core \citep{Larson1969}.  
The formation of the stellar core occurs just
a few years after the onset of the second collapse, during which the maximum density 
increases from $\sim 10^{-8}$ to $\gsim 0.1$ g~cm$^{-3}$ and the maximum temperature
increases from $\approx 2000$ to $>60,000$~K.  The stellar core is formed with a mass
of $M_{\rm sc} \approx 1.5$~M$_{\rm J}$ and a radius of $R_{\rm sc} \approx 2$~R$_\odot$.  Without
rotation, the stellar core accretes the remnant of the first core in which it is embedded
in $\approx 10$~years and then accretes the envelope \citep{Larson1969}, though with three-dimensional
calculations we only follow the calculations for $\approx 50-100$ years after stellar core formation.

\begin{figure}
\centering \vspace{-0.0cm}
    \includegraphics[width=17cm]{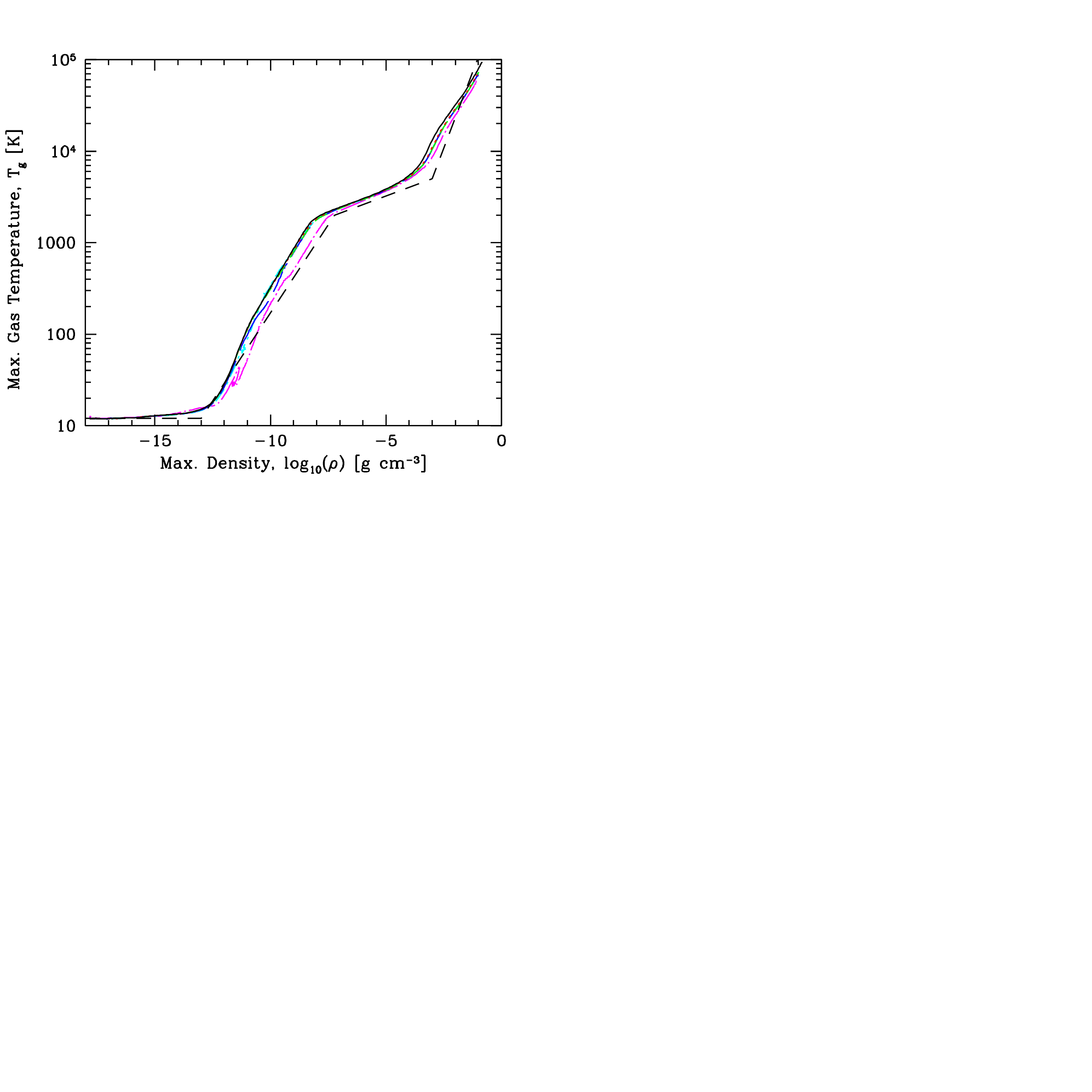} \vspace{-10.0cm}
\caption{The evolution of the maximum gas temperature versus maximum density for the radiation hydrodynamical collapse of molecular cloud cores with $\beta=0$ (solid black line), $5\times 10^{-4}$ (dotted red curve), 0.001 (short-dashed green curve), 0.005 (long-dashed blue curve), 0.01 (dot-short dashed cyan curve), 0.04 (dot-long-dashed magenta curve).  The barotropic equation of state is given by the black short-dashed lines.  The axisymmetric ($\beta=0$) case is always the hottest at a given maximum density because the radiation is most effectively trapped.  Conversely, the calculation that is the coolest has the highest rotation rate ($\beta=0.04$).  
Each calculation was performed with $10^6$ SPH particles.}
\label{temp_vs_density}
\end{figure}

Although these stages are qualitatively the same in the barotropic calcuations, there is a much greater difference in the evolution of the first core between 
the barotropic and radiation hydrodynamical calculations than for the phase of the 
collapse prior to first core formation.  To make this clear, in Fig.~\ref{first_core_time} the
time evolution of maximum density and maximum temperature during the calculations is
replotted with $t=0$ set to the time of stellar core formation (defined as the time when the maximum
density reaches $10^{-3}$~g~cm$^{-3}$).  This allows us to clearly see the amount of
time spent between first core formation and stellar core formation in each of the calculations.
The barotropic results are plotted using dashed lines, while the solid lines give the radiation
hydrodynamical results.  The evolution time of the first core or pre-stellar disc is longer using 
radiation hydrodynamics than using the barotropic equation of state in all cases, by factors of $1.5-3$ 
except for the most rapidly-rotating case.
This is because the barotropic equation of state consistently underestimates the temperature 
of the gas, providing less pressure support to the gas and, thus, allowing it to enter the second
collapse phase earlier.  The consistent temperature underestimate can be seen 
clearly in Fig.~\ref{temp_vs_density}
which plots the maximum temperature versus maximum density for each of the radiation
hydrodynamical calculations and compares this to the barotropic equation of state 
(dashed black solid line).  The lines from the radiation hydrodynamical calculations
almost lie on top of one another, but from densities of $10^{-11}$ to $10^{-8}$~g~cm$^{-3}$
the barotropic equation of state underestimates the maximum temperature by about a factor
of two.  The only exception to this is the most rapidly-rotating calculation with $\beta=0.04$.
Here the maximum temperature at a given density is lower than in the other radiation hydrodynamical
calculations and more similar to the barotropic equation of state.  This is because in this case the
first core is actually a torus and, therefore, the gas can cool more effectively.  Returning to 
Fig.~\ref{first_core_time}, we also see that it is the $\beta=0.04$ case (rightmost panels) 
where the timescales between first and stellar core formation are most similar for the barotropic
and radiation hydrodynamical calculations (differing only by about 15\% rather than a factor of $1.5-3$).

\begin{figure}
\centering \vspace{-0.0cm}
    \includegraphics[width=17cm]{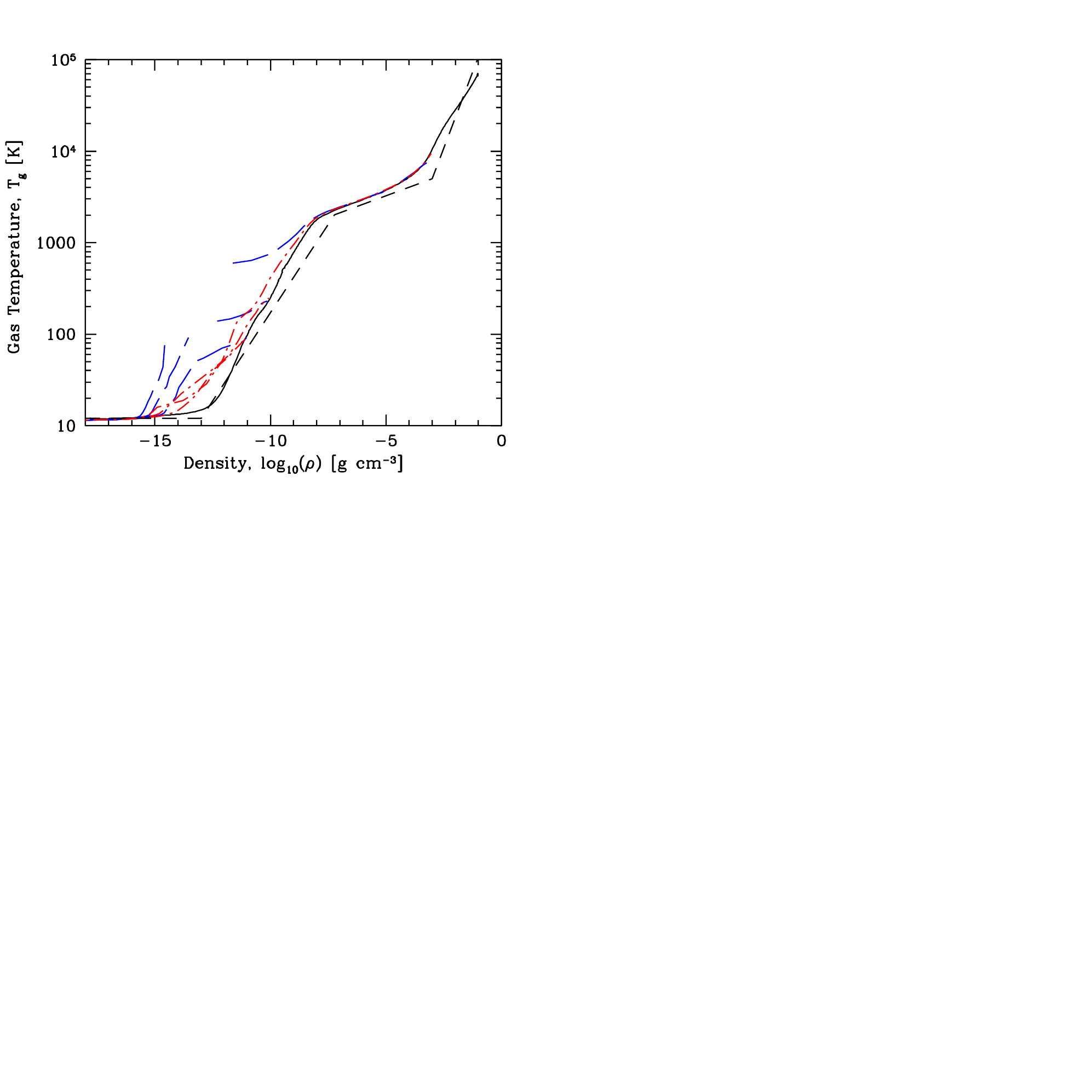} \vspace{-10.0cm}
\caption{For the radiation hydrodynamical collapse of the molecular cloud core with $\beta = 0.005$, we give the evolution of the maximum gas temperature versus maximum density during the collapse (solid black curve) and snapshots of gas temperature versus density along the rotation ($z$) axis (long-dashed blue curves) and perpendicular to the rotation axis (dot-dashed red curves) at three different times.  Note that the long-dashed blue lines at higher temperatures (later times) have a break in them.  This is because there is not enough gas along the rotation axis at these densities to determine an accurate temperature.  These densities occur within the accretion shock onto the first core. Although the evolution of the maximum temperature versus maximum density is similar to that given by the barotropic equation of state (black short-dashed lines), the gas temperature surrounding the protostar tends to be much hotter at a given density than predicted by the barotropic equation of state, particularly along the rotation axis where the temperatures can be more than an order of magnitude hotter.
This calculation was performed with $3 \times 10^6$ SPH particles.}
\label{temp_vs_density_snapshot}
\end{figure}

One possible reason that the radiation hydrodynamical calculations give greater temperatures
during the first core phase than the barotropic calculations is that there may be considerable shock heating which is
not taken into account with a barotropic equation of state.  However,
computing the entropy of the gas before the first core forms and in the bulk of the first core
as it forms and evolves, we find it monotonically decreases in the radiation
hydrodynamical calculations.  The rate of
decrease is rapid before the first core forms (when it is cooling rapidly and almost isothermal), 
but even after the first core forms the gas is loosing energy due to radiation.  Only at the surface 
of the first core (around the accretion shock) does the entropy briefly increase before it radiatively
cools.  This is consistent with the recent calculations of \cite{Commerconetal2011b} who find that 
essentially all of the energy liberated in the accretion shock is radiated away.

Instead, the reason the barotropic equation of state consistently underestimates the temperature
in this density range is due to the approximation that $\eta=7/5$ in this part of the evolution.
In fact, molecular hydrogen (the dominant constituent) has a ratio of specific
heat capacities of $\gamma=5/3$
until it reaches $\sim 100$~K.  Only then are the rotational degrees of freedom, which lower
$\gamma$ to 7/5, excited.  Again, this is apparent in Fig.~\ref{temp_vs_density} where it can
be seen that the lines from the radiation hydrodynamical calculations are steeper than the 
barotropic line in the temperature range $\approx 20-150$.  Above this temperature, the lines
are almost parallel (i.e.\ $\gamma \approx 7/5$ for both), but the temperature offset
(that originates in the $20-150$~K range) persists.  The barotropic equation of
state could be improved for this part of the evolution by including a smooth transition from
$\gamma=1$ to $5/3$ and then a transition from $5/3$ to $7/5$. 
However, this would still not
capture all of the detail present in the radiation hydrodynamical equations (see below).

\subsection{Rotationally-unstable first hydrostatic cores}
\label{rapidly_rotating}

If the first core is rotating rapidly enough that its own value of $\beta > 0.274$, the
core is dynamically unstable to the growth of non-axisymmetric structure \citep{Bate1998,Machidaetal2005,SaiTom2006,SaiTomMat2008,Bate2010, MacInuMat2010,Tomidaetal2010b,SaiTom2011, MacMat2011}.
For the particular initial conditions used here, this occurs for the $\beta=0.001-0.01$ cases 
(see Figs.~\ref{images_D_barotropic} and \ref{images_xyD}).  The torus that forms in the 
$\beta=0.04$ calculations is also dynamically unstable, but this case is somewhat different and
will be discussed further below.

In each of the $\beta=0.001-0.005$ cases, the first core begins as an axisymmetric flattened 
pre-stellar disc, but after several rotations it develops a bar-mode (e.g. the second panel of the third row in 
each of Figures \ref{images_D_barotropic} and \ref{images_xyD}).  The ends of the bar 
subsequently lag behind and the bar winds up to produce a 
spiral structure.
Spiral structure removes angular momentum from the inner parts of the first
core via gravitational torques \citep{Bate1998}, the effect of which can be seen in the 
evolution of density and temperature in
Fig.~\ref{first_core_time}.  For example, in the $\beta=0.005$ case, the slow increases in
central density and temperature after first core formation suddenly accelerate with the onset of the 
spiral structure.  This occurs at about 1300~yrs before stellar core formation 
for the radiation hydrodynamical calculation with
$\beta=0.005$ and about 600~yrs before stellar core formation for the barotropic calculation.  
Similar accelerations
are seen for both the $\beta=0.01$ cases and also for the barotropic $\beta=0.001$ case.  
Thus, the removal of angular momentum from the
central regions of the core substantially accelerates the evolution of the first core towards 
the second collapse, which would take much longer to reach without the
angular momentum redistribution (i.e.\ due to accretion and radiative cooling alone). 

\begin{figure*}
\centering \vspace{-0.5cm}
    \includegraphics[width=19.0cm]{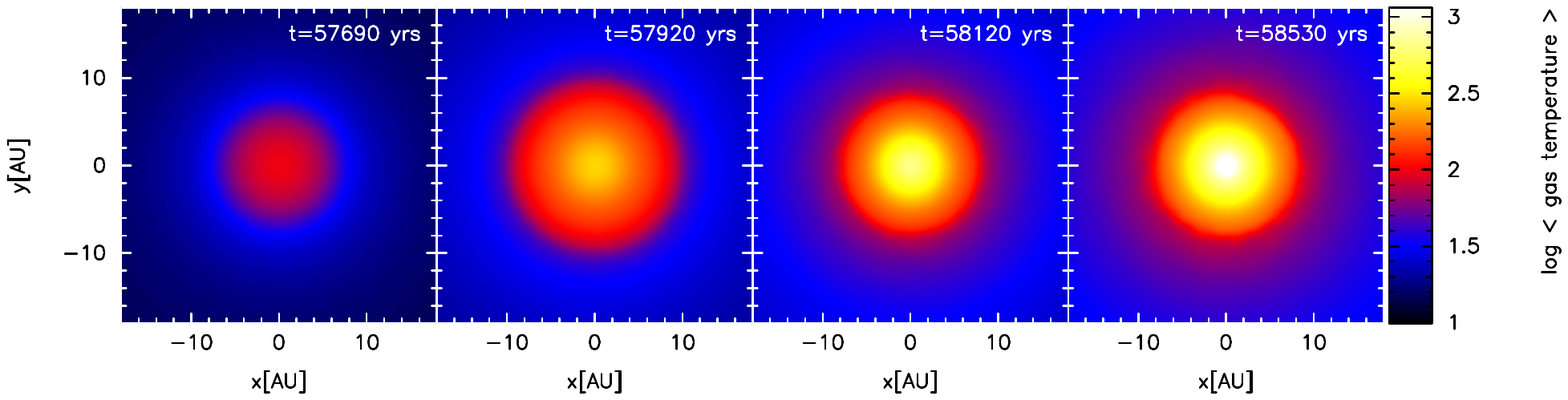}\vspace{-20.7cm}
    \includegraphics[width=19.0cm]{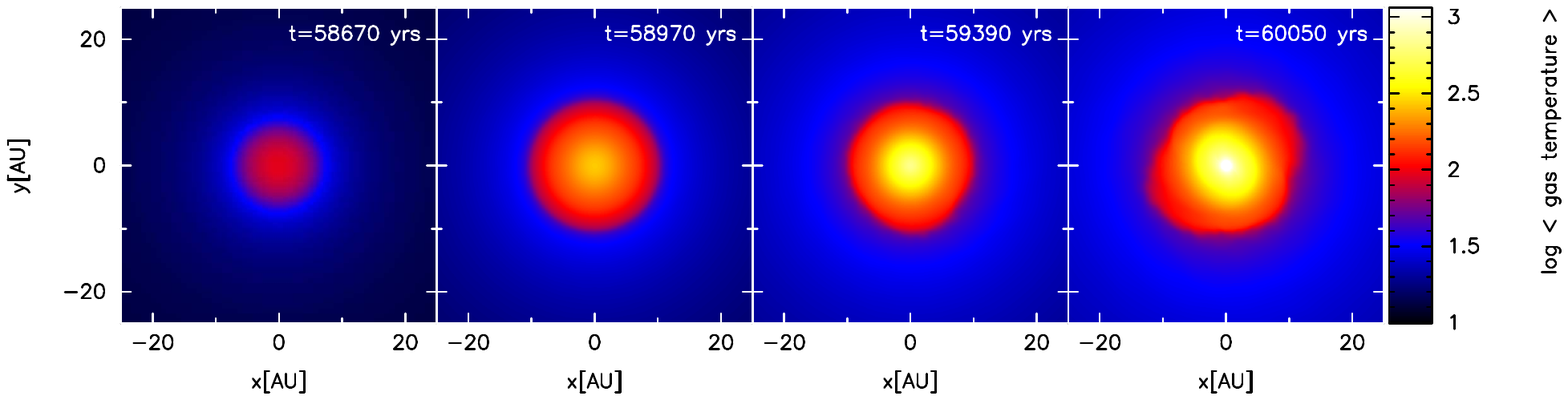}\vspace{-20.7cm}
    \includegraphics[width=19.0cm]{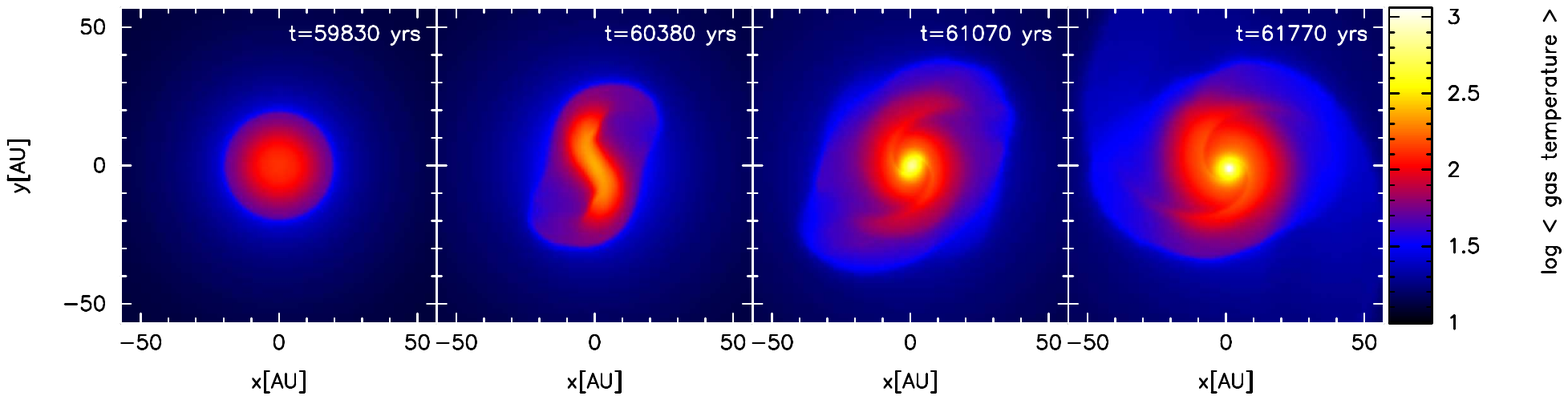}\vspace{-20.7cm}
    \includegraphics[width=19.0cm]{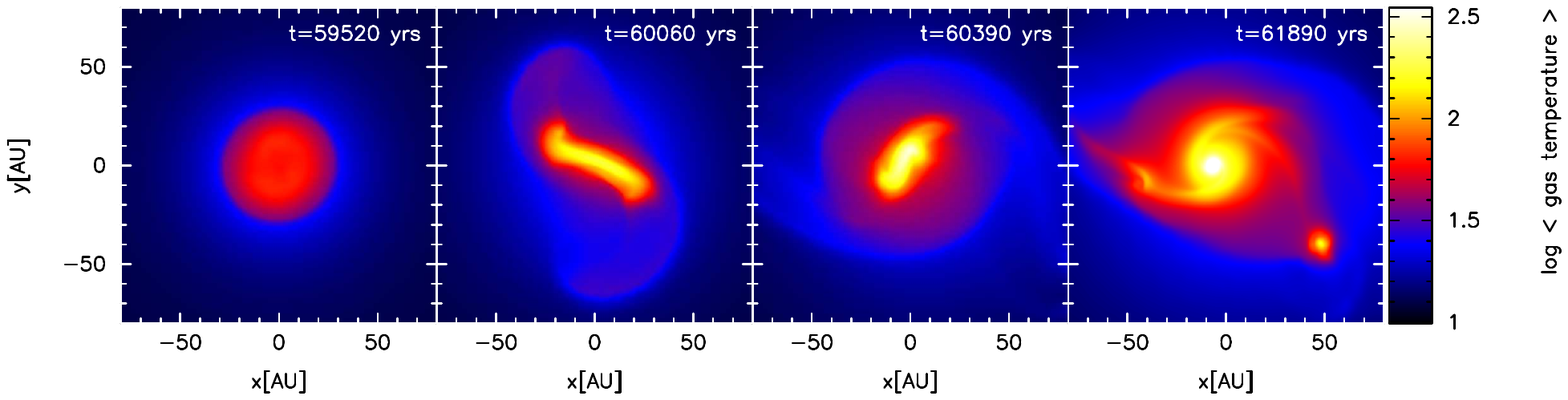}\vspace{-20.7cm}
    \includegraphics[width=19.0cm]{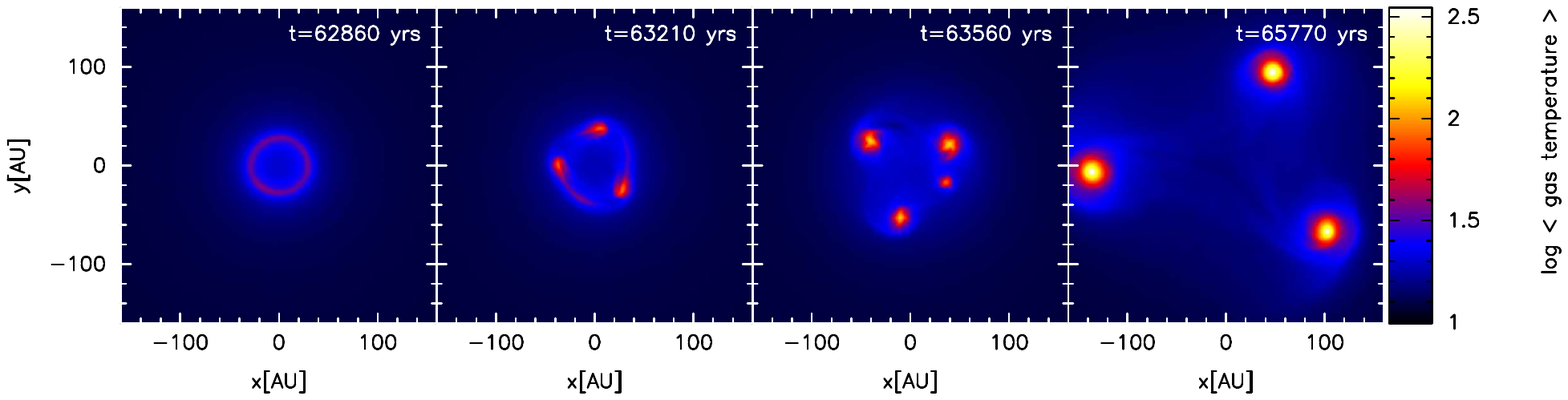}\vspace{-19.5cm}
\caption{Snapshots of the density-weighted temperature viewed parallel to the rotation axis during the evolution of the radiation hydrodynamical calculations of the collapse of molecular cloud cores with different initial rotation rates.  From top to bottom, the different rows are for cloud cores with $\beta=5\times 10^{-4},0.001,0.005,0.01,0.04$. Note that the spatial scale is different for each row, with each panel measuring $1600 \sqrt{\beta}$ AU across (i.e. from 36 to 320 AU).   The temperature sale differs between the top three rows and the lower two rows in order to enhance the low-temperature structure in the cases with greater rotation.  The free-fall time of the initial cloud core, $t_{\rm ff}=1.8\times 10^{12}$~s (56,500 yrs).  
Each calculation was performed with $10^6$ SPH particles, except for the $\beta=0.001$ and $\beta=0.005$ cases which used $3 \times 10^6$ SPH particles.  The evolution is almost identical when using $10^6$ or $3 \times 10^6$ particles, but the latter are slightly more detailed (see Appendix A).
}
\label{images_xyT}
\end{figure*}

In Figs.~\ref{images_D_barotropic} and \ref{images_xyD}, the development of the spiral 
structure and the subsequent concentration of material towards the centre of the core due to
the redistribution of angular momentum is clearly visible for the barotropic cases with 
$\beta=0.001-0.01$ and the radiation hydrodynamical cases with $\beta=0.005-0.01$.
In Fig.~\ref{images_xyT} we also provide the density-weighted temperature, for comparison
with the column density in Fig.~\ref{images_xyD}.  As gas is concentrated to the 
centre of the cores its temperature greatly increases.  It is also apparent that the spiral
shocks in the discs are hotter than the rest of the discs.

The barotropic and radiation hydrodynamical evolutions are qualitatively similar in that both display
the progression from an axisymmetric core, to bar instability, to a torus as the initial rotation rate of
the molecular cloud core is increased.  However, radiation hydrodynamical calculations are
more resistant to the bar instability than the barotropic calculations.  This is directly attributable to
the difference in the thermal evolution of the gas that was discussed above.  Since the gas is 
somewhat hotter in the radiation hydrodynamical calculations, the first core is larger (more `puffy')
and has a lower value of $\beta$ for a given $\beta$ value of the initial molecular cloud core.
\cite{Tomidaetal2010a} also noticed that the first cores in their radiation magnetohydrodynamical calculations 
had higher entropies and larger sizes than when they used a barotropic equation of state.
It should be noted, however, that with radiation hydrodynamics there is no one-to-one relation
between temperature and density \citep[see also][]{Bossetal2000,WhiBat2006,Tomidaetal2010a}.  This is illustrated in
Fig.~\ref{temp_vs_density_snapshot}, where we plot temperature versus density from various
snapshots from the $\beta=0.005$ calculation and compare this to both the run of 
maximum temperature versus maximum density during the calculation and the 
barotropic equation of state.  The red dot-dashed lines are for cuts perpendicular to the rotation
axis along the $x$-axis (i.e. in the plane of the disc), while the blue long-dashed lines
display the temperature along the rotation axis.  Not only is the run of maximum temperature versus
maximum density always greater than or equal to the temperature from the 
barotropic equation of state, but the gas both in the midplane of the disc and along the rotation
axis at the same density is even hotter.  This is particularly true of the gas along the rotation axis,
which is heated by the accretion shock at the surface of the first core.  \cite{Tomidaetal2010a}
give a very similar plot for a snapshot from one of their simulations and find very similar
behaviour.

These hotter temperatures in the radiation hydrodynamical calculations mean that while the 
transition from rotationally stable to dynamically unstable first core occurs in the range 
$\beta=5\times 10^{-4}-10^{-3}$
using the barotropic equation of state, the transition occurs at $\beta\approx 0.001$ with
radiation hydrodynamics (this case undergoes an extremely weak instability, just visible in
the fourth panel of the second row of Fig.~\ref{images_xyD}).  Similarly, while the barotropic 
calculation with $\beta=0.01$ manages to produce a torus-shaped first core, the radiation
hydrodynamical calculation with $\beta=0.01$ is still definitely disc shaped rather than torus shaped.

As mentioned above, for the highest initial rotation rates ($\beta=0.04$) the first core is actually
a torus or ring-like structure \citep[e.g.][]{NorWil1978,ChaWhi2003,Machidaetal2005}.  Such a configuration is
highly unstable to non-axisymmetric perturbations and, indeed, as is clearly visible in 
Figs.~\ref{images_D_barotropic} and \ref{images_xyD} the rings rapidly fragment into four 
objects.  Such a configuration is highly chaotic, and symmetry is broken quickly (due to truncation error and
the use of a tree-structure to calculate gravity in the calculations).  In the radiation hydrodynamical
calculation, two of the fragments merge to produce a triple system, while in the barotropic calculation
all four fragments survive (at least until the calculation was stopped).  Each of the fragments follows
its own evolution toward the second collapse phase and stellar core formation.  The calculations
were stopped soon after the first fragment in each calculation produced a stellar core.

Finally, as mentioned above, first cores may evolve into pre-stellar discs 
with radii ranging from $\approx 5$ to $\gsim 100$~AU 
before a stellar core forms (Fig.~\ref{images_xyD}).  Those with radii greater 
than $\approx 10$~AU are produced due to the angular momentum transport that occurs
during the dynamical rotational instability.  A further consequence is that all star formation
should go through at least a brief phase when the disc mass is greater than the stellar mass.
Such discs are gravitationally unstable and may evolve through spiral density waves
(e.g. the third row of Figs.~\ref{images_xyD} and ~\ref{images_xyT}) 
and/or fragmentation (e.g. the fourth rows of Fig.~\ref{images_xyD} and ~\ref{images_xyT}).  
This is discussed further in Section \ref{discussion}.

In conclusion, in switching from the simple barotropic equation of state to a realistic equation of state
with radiation hydrodynamics,
the qualitative evolution of the first core and its dependence on the initial rotation rate of
the molecular cloud core is identical.  However, the gas temperature in the more realistic 
calculations tends to be hotter and is not only a function of density: for example, the gas is significantly 
hotter in the accretion shock at the surface of the first cores.   Quantitatively, 
the higher temperatures mean that the critical values of the initial molecular cloud core rotation
rates required for bar instability of the first core, or the transition to a torus geometry is somewhat 
higher.  The value of $\beta$ must be $\approx 50$\% greater with radiation hydrodynamics,
which translates into a rotation rate which is $\approx 25$\% greater (since $\beta \propto \Omega^2$,
where $\Omega$ is the angular frequency of the molecular cloud core).

\begin{figure*}
\centering \vspace{-0.5cm}
    \includegraphics[width=18.0cm]{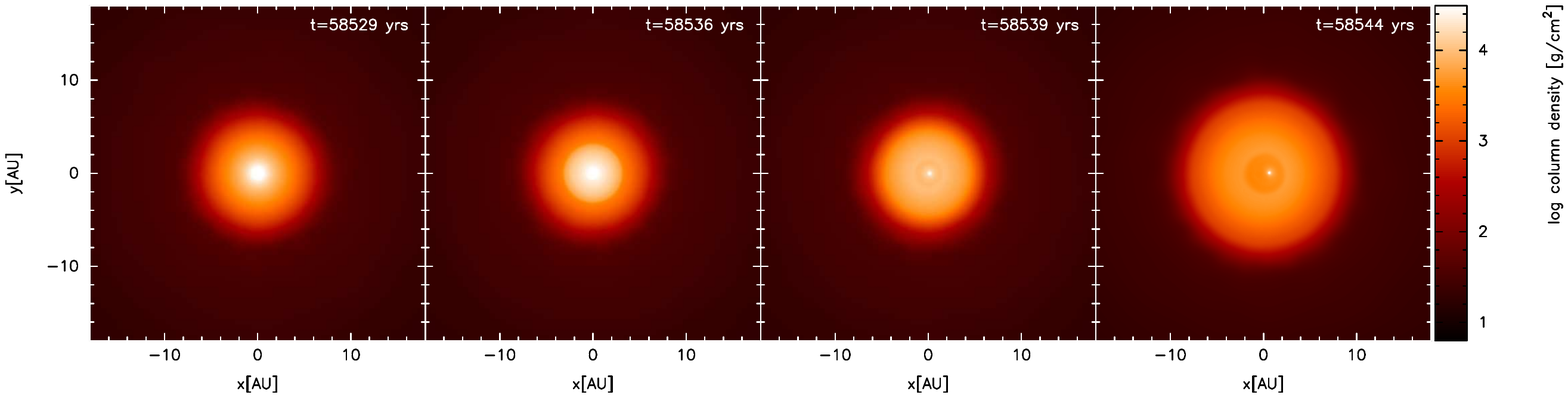}\vspace{-9.9cm}
    \includegraphics[width=18.0cm]{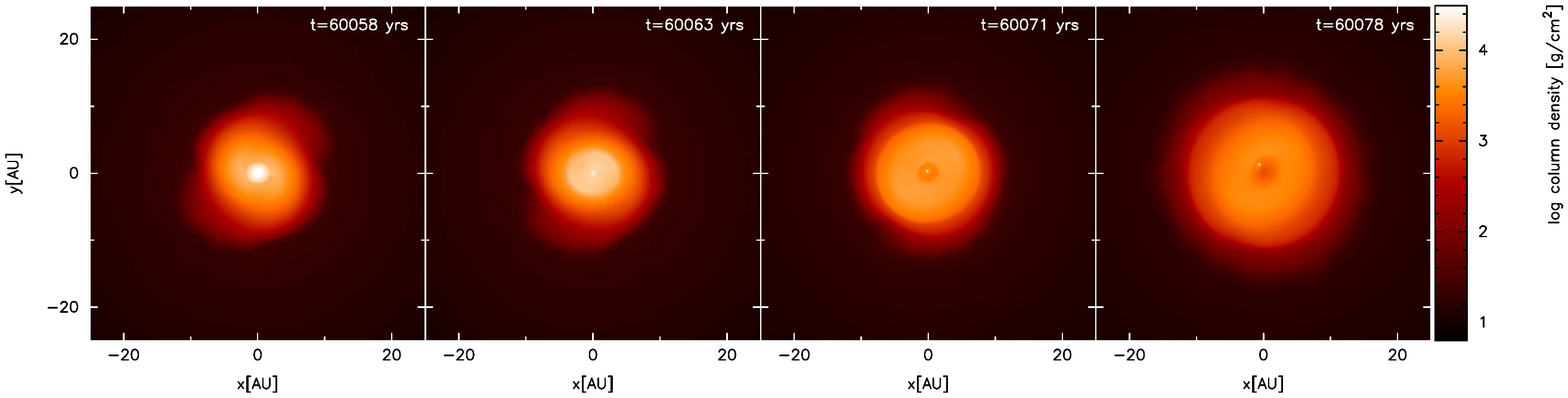}\vspace{-9.9cm}
    \includegraphics[width=18.0cm]{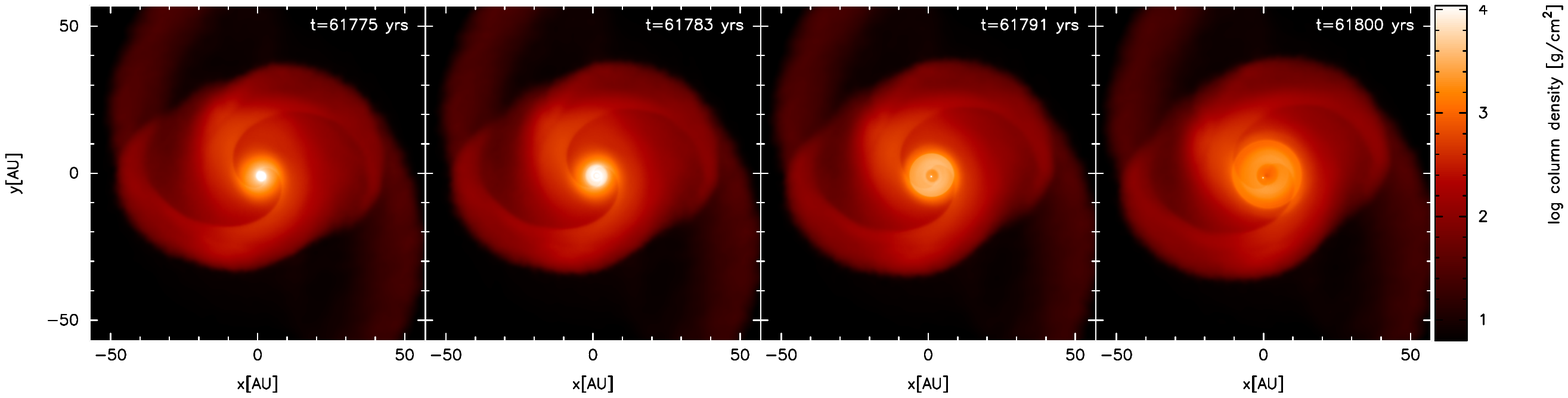}\vspace{-9.9cm}
    \includegraphics[width=18.0cm]{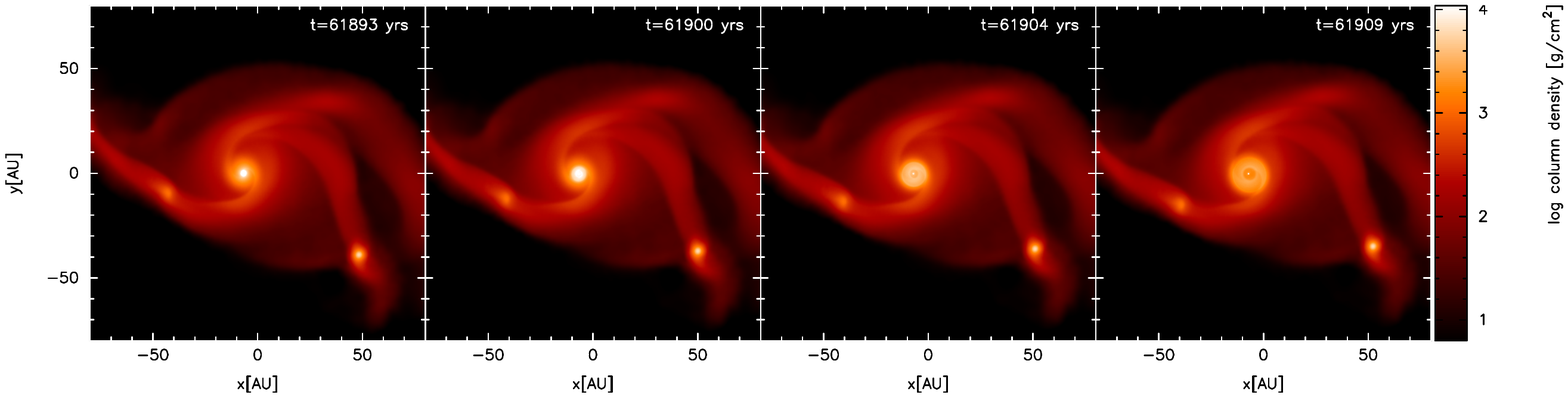}
\vspace{-9.1cm}
\caption{Snapshots of the column density viewed parallel to the rotation axis during the evolution of the radiation hydrodynamical calculations of the collapse of molecular cloud cores with different initial rotation rates.  The shockwaves propagating outwards through the discs following stellar core formation are clearly visible. From top to bottom, the different rows are for cloud cores with $\beta=5\times 10^{-4},0.001,0.005,0.01,0.04$. Note that the spatial scale is different for each row, with each panel measuring $1600 \sqrt{\beta}$ AU across (i.e. from 36 to 320 AU).  The free-fall time of the initial cloud core, $t_{\rm ff}=1.8\times 10^{12}$~s (56,500 yrs).  
Each calculation was performed with $10^6$ SPH particles, except for the $\beta=0.001$ and $\beta=0.005$ cases which used $3 \times 10^6$ SPH particles.  The evolution is almost identical when using $10^6$ or $3 \times 10^6$ particles, but the latter are slightly more detailed (see Appendix A).
}
\label{images_xyD_OUTFLOW}
\end{figure*}

\begin{figure*}
\centering \vspace{-0.5cm}
    \includegraphics[width=18.0cm]{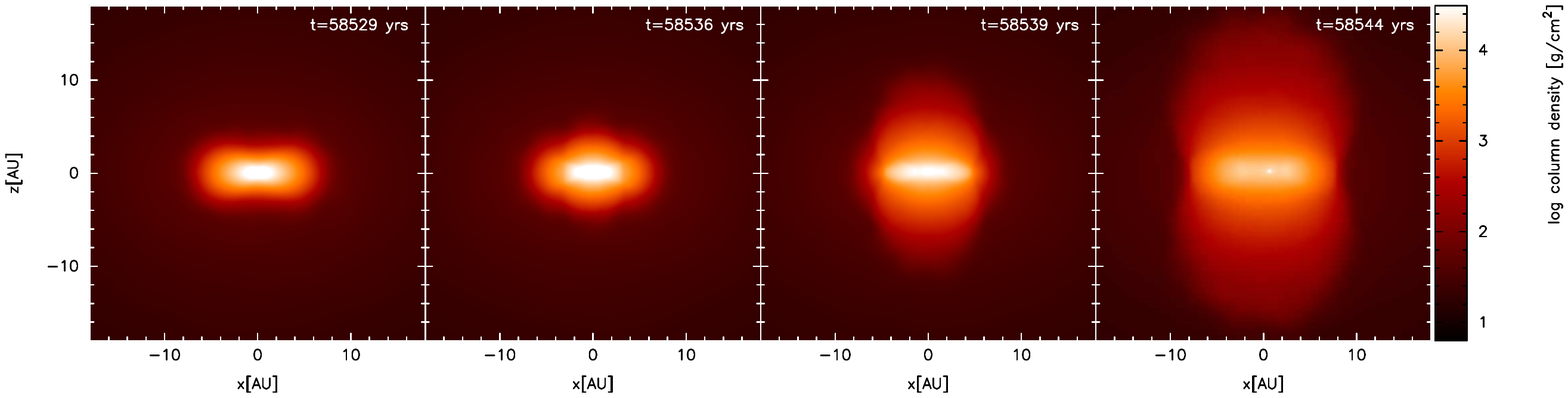}\vspace{-9.9cm}
    \includegraphics[width=18.0cm]{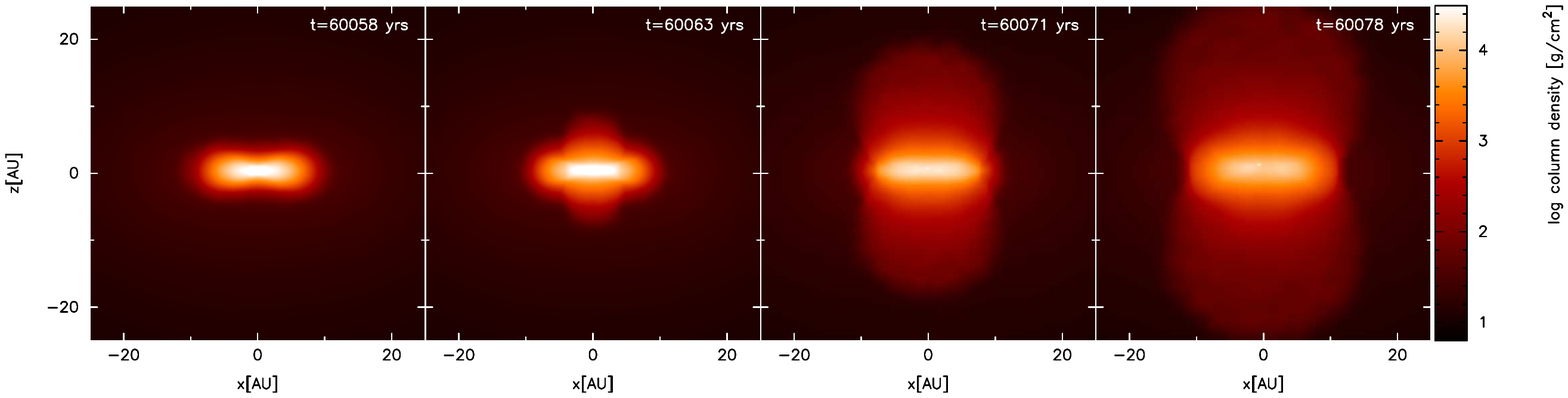}\vspace{-9.9cm}
    \includegraphics[width=18.0cm]{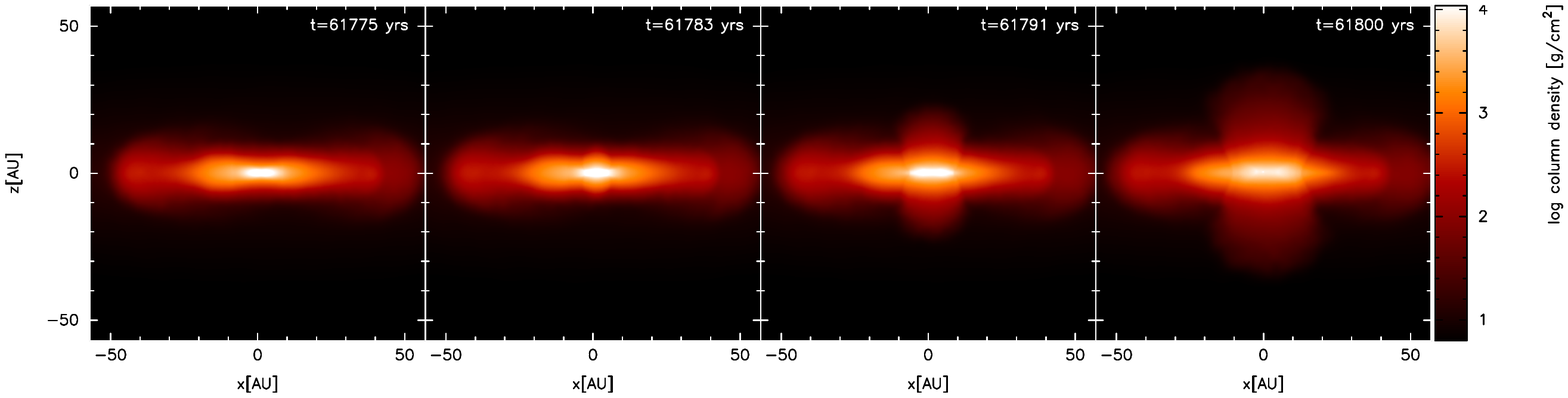}\vspace{-9.9cm}
    \includegraphics[width=18.0cm]{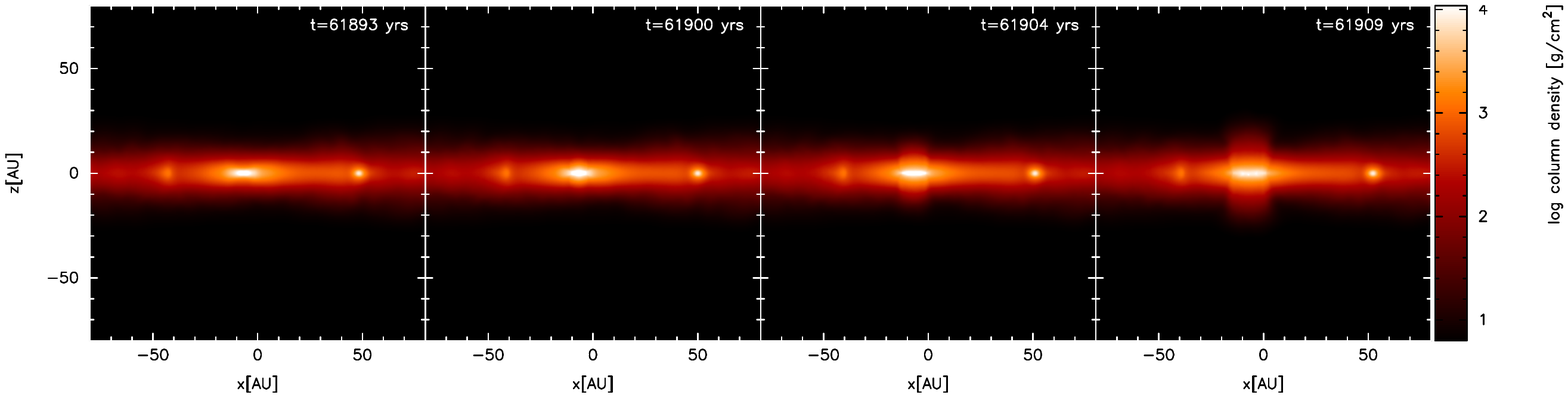}
\vspace{-9.1cm}
\caption{Snapshots of the column density viewed perpendicular to the rotation axis during the evolution of the radiation hydrodynamical calculations of the collapse of molecular cloud cores with different initial rotation rates.  The bipolar outflows, launched after stellar core formation, are clearly visible. From top to bottom, the different rows are for cloud cores with $\beta=5\times 10^{-4},0.001,0.005,0.01,0.04$. Note that the spatial scale is different for each row, with each panel measuring $1600 \sqrt{\beta}$ AU across (i.e. 36, 50, 114, 160, or 320 AU).  The free-fall time of the initial cloud core, $t_{\rm ff}=1.8\times 10^{12}$~s (56,500 yrs).  
Each calculation was performed with $10^6$ SPH particles, except for the $\beta=0.001$ and $\beta=0.005$ cases which used $3 \times 10^6$ SPH particles.  The evolution is almost identical when using $10^6$ or $3 \times 10^6$ particles, but the latter are slightly more detailed (see Appendix A).
}
\label{images_xzD_OUTFLOW}
\end{figure*}

\begin{figure*}
\centering \vspace{-0.5cm}
    \includegraphics[width=17.0cm]{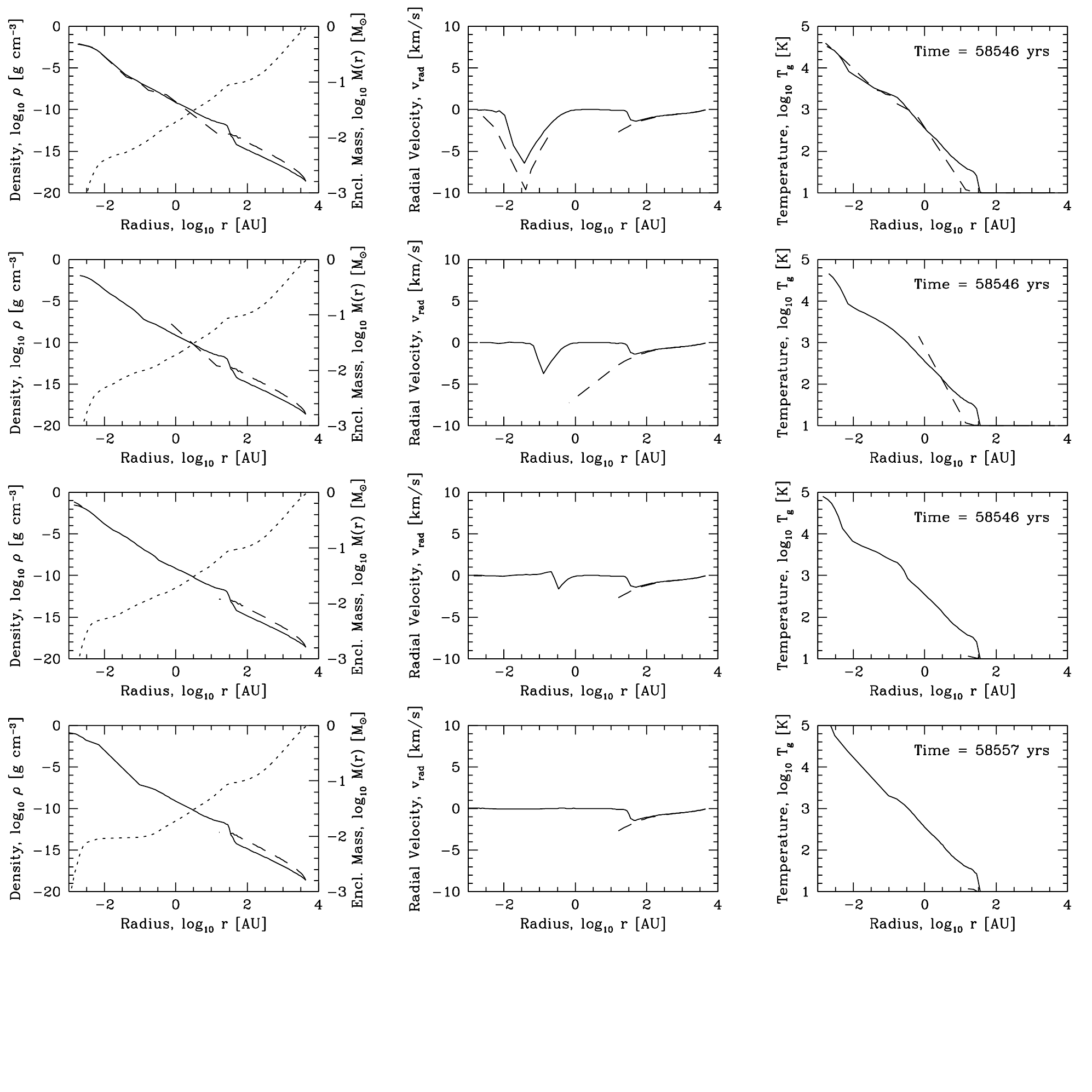}\vspace{-2.5cm}
\caption{The evolution of the barotropic $\beta=0.005$ calculation of the collapse of a molecular cloud core following stellar core formation.  Each of the four rows shows the state of the protostar at different time.  The left panels provide the radial density profile perpendicular to the rotation axis (solid line, averaged in azimuth), the density profile along the rotation axis (dashed line), and the cumulative mass profile (dotted line).  The centre panels give the radial velocity profiles perpendicular to (solid line) or along (dashed line) the rotation axis.  The right panels show the radial temperature profiles perpendicular to (solid line) or along (dashed line) the rotation axis.  Breaks in the lines occur when there is not sufficient gas resolution to define an accurate measurement.}
\label{lines_baro}
\end{figure*}

\begin{figure*}
\centering \vspace{-0.5cm}
    \includegraphics[width=17.0cm]{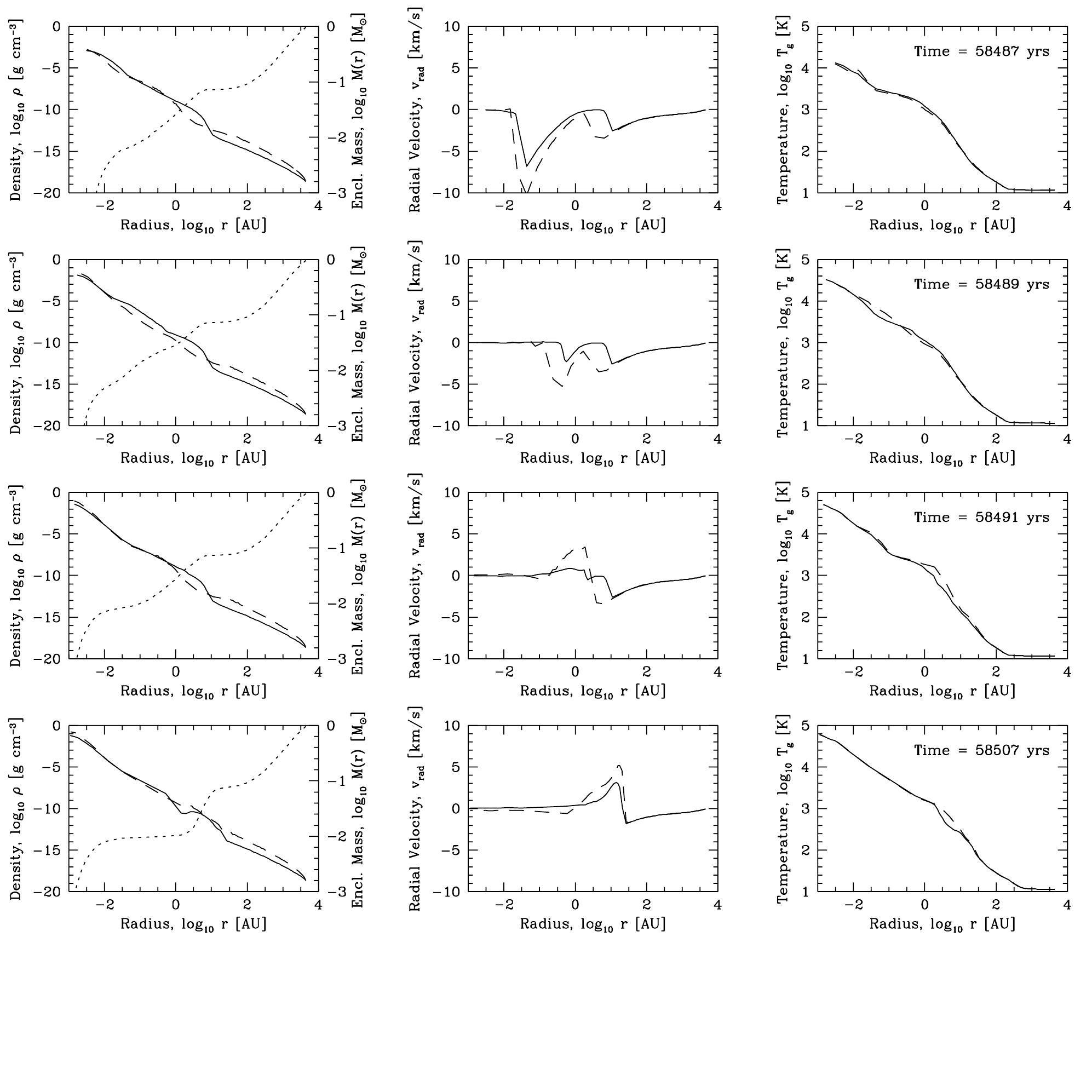}\vspace{-2.5cm}
\caption{The evolution of the radiation hydrodynamical $\beta=5\times 10^{-4}$ calculation of the collapse of a molecular cloud core following stellar core formation.  Each of the four rows shows the state of the protostar at different time.  The left panels provide the radial density profile perpendicular to the rotation axis (solid line, averaged in azimuth), the density profile along the rotation axis (dashed line), and the cumulative mass profile (dotted line).  The centre panels give the radial velocity profiles perpendicular to (solid line) or along (dashed line) the rotation axis.  The right panels show the radial temperature profiles perpendicular to (solid line) or along (dashed line) the rotation axis.}
\label{lines_beta0_0005}
\end{figure*}

\begin{figure*}
\centering \vspace{-0.5cm}
    \includegraphics[width=17.0cm]{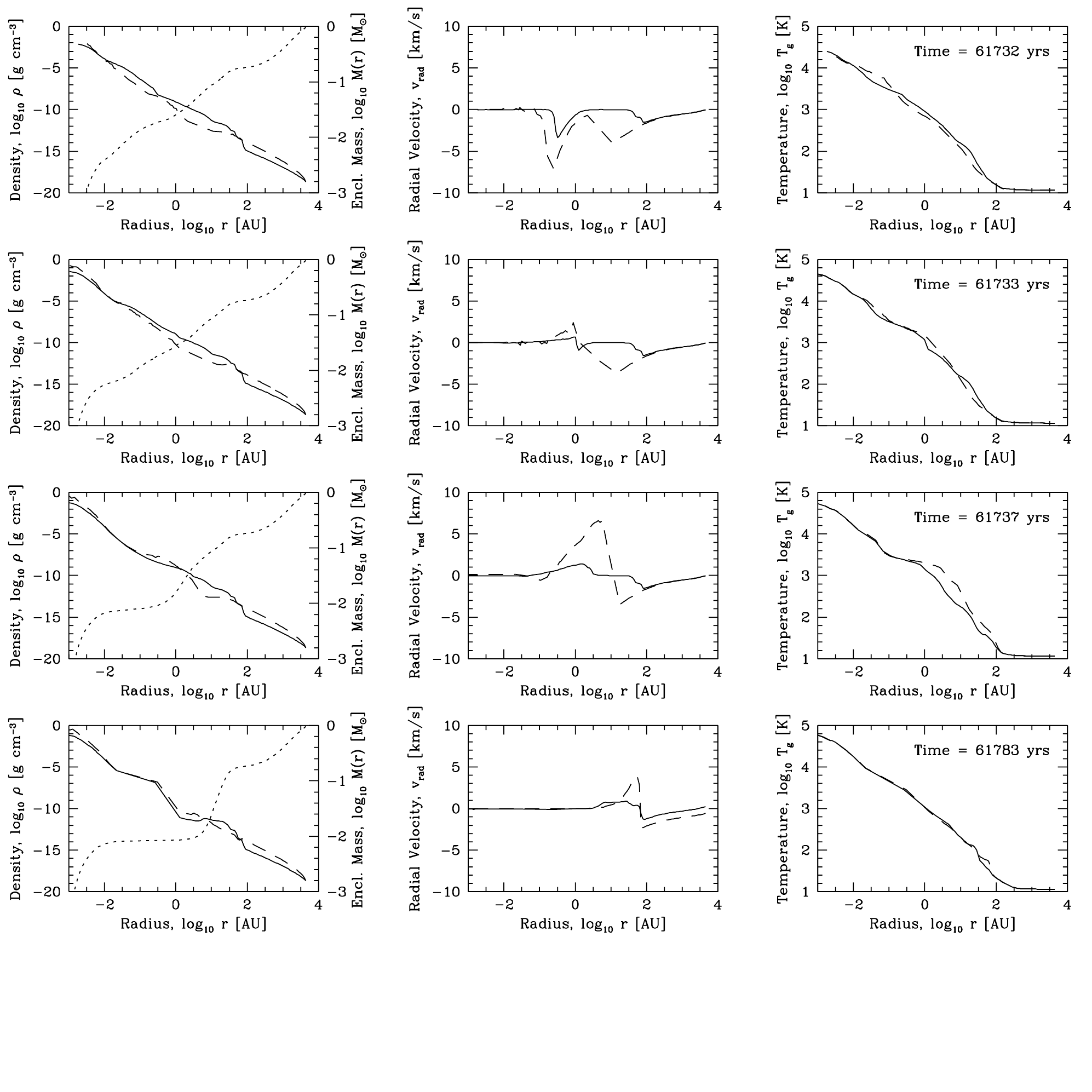}\vspace{-2.5cm}
\caption{The evolution of the radiation hydrodynamical $\beta=0.005$ calculation of the collapse of a molecular cloud core following stellar core formation.  Each of the four rows shows the state of the protostar at different time.  The left panels provide the radial density profile perpendicular to the rotation axis (solid line, averaged in azimuth), the density profile along the rotation axis (dashed line), and the cumulative mass profile (dotted line).  The centre panels give the radial velocity profiles perpendicular to (solid line) or along (dashed line) the rotation axis.  The right panels show the radial temperature profiles perpendicular to (solid line) or along (dashed line) the rotation axis.}
\label{lines_beta0_005}
\end{figure*}

\subsection{The effect of stellar core formation}
\label{stellar_core_formation}

Although the evolution of the first core or pre-stellar disc is qualitatively the same when computed with a
barotropic equation of state or radiation hydrodynamics, as \cite{Bate2010} showed 
the evolution subsequent to the 
formation of the stellar core is {\it qualitatively different}.
When using a barotropic equation of state, the formation of the stellar core deep within
the optically-thick disc has no effect on the temperature of the
gas further out in the disc because its temperature is set purely according to its density.
However, with radiation hydrodynamics, the situation is completely different.

\begin{figure*}
\centering \vspace{-0.0cm}
    \includegraphics[width=8.5cm]{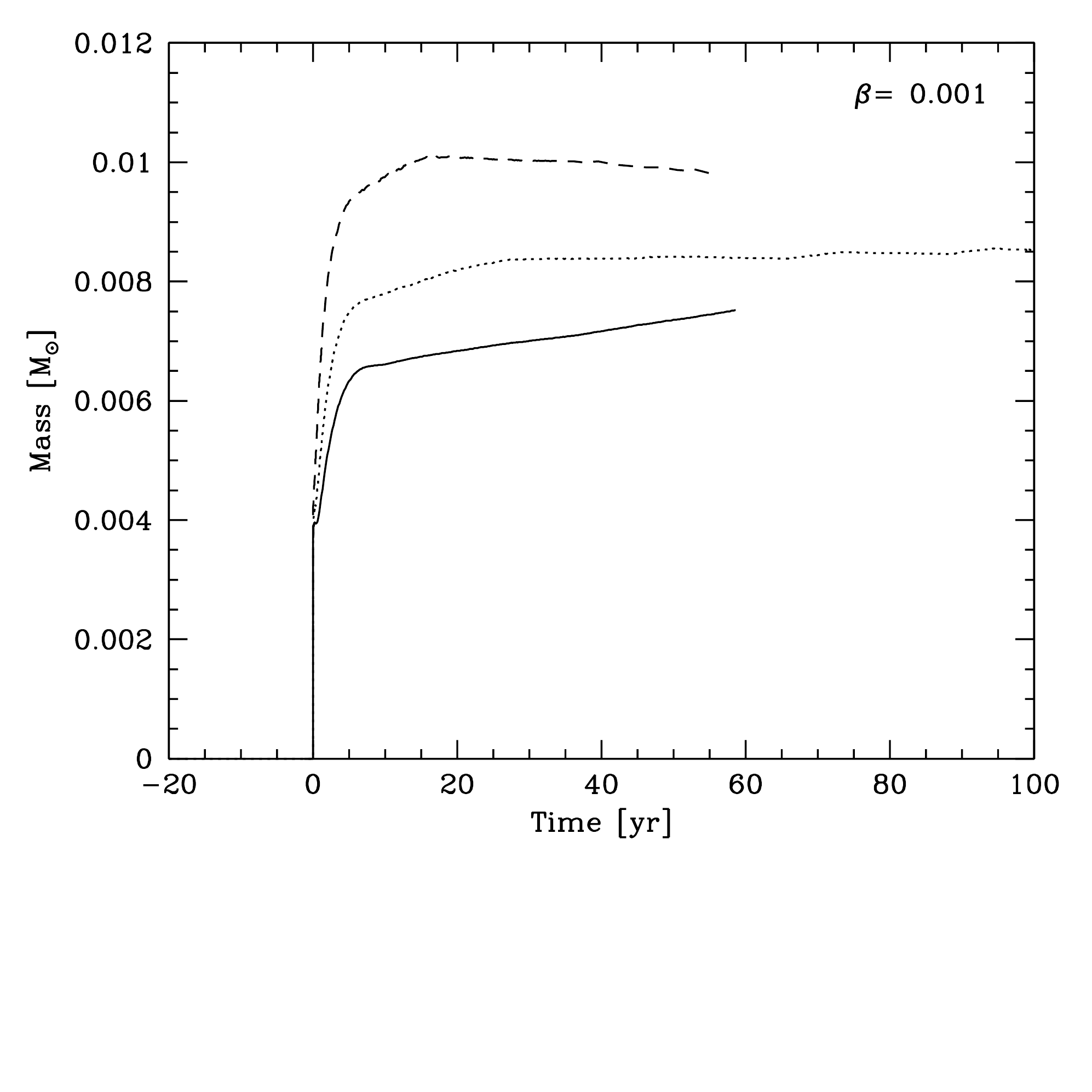}
    \includegraphics[width=8.5cm]{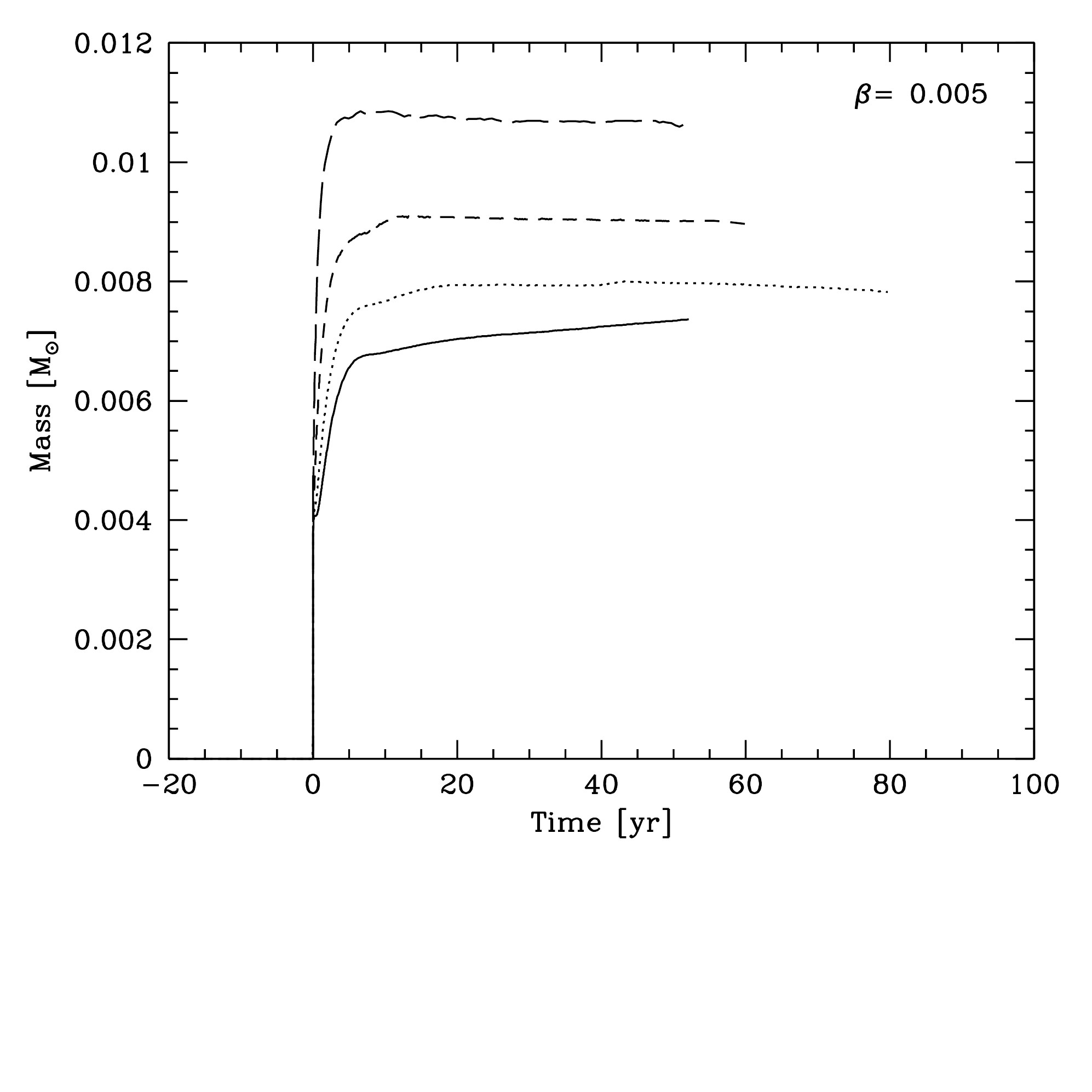}\vspace{-2.0cm}
\caption{The mass of the stellar core (gas with density $>10^{-4}$~g~cm$^{-3}$) measured from the time of its formation.  The two panels shows the results for the radiation hydrodynamical $\beta=0.001$ (left) and $\beta=0.005$ (right) molecular cloud cores performed using $10^5$ (long-dashed line), $3\times 10^5$ (short-dashed line), $10^6$ (dotted line), and $3\times 10^6$ (solid line) particles.  With low resolutions the radiative feedback stops accretion onto the stellar cores, but for the highest resolutions the accretion continues at a low level ($1-2 \times 10^{-5}$~M$_\odot$~yr$^{-1}$). }
\label{convergence}
\end{figure*}

When the second collapse occurs and produces the stellar core, the gravitational
potential energy that is released is $\sim GM_{\rm sc}^2/R_{\rm sc} = 4 \times 10^{42}$~erg.
Since the stellar core is in virial equilibrium, approximately half of this energy is
radiated away.  Moreover, the stellar core rapidly begins to accrete from the
first core, reaching a mass of $\approx 6$~M$_{\rm J}$ in only a few years (i.e. more 
quickly than the dynamical timescale of the large-scale disc; see below).  This increases the total 
energy released to $\approx 3 \times 10^{43}$~erg.
\cite{Bate2010} compared this energy with the binding energy of the disc in 
the $\beta=0.005$ calculation. Just before the onset of the second collapse, the binding 
energy of the pre-stellar disc is only $4 \times 10^{43}$~erg 
(estimated as $\sim GM_{\rm d}^2/R_{\rm d}$ with  
$M_{\rm d} \approx 0.18~{\rm M}_\odot$ and taking a mean `spherical' radius of $R_{\rm d} \approx 15$~AU).
Thus, when the second collapse occurs, the disc suddenly finds itself irradiated from the 
inside by an energy source emitting a substantial fraction of the binding energy of the disc itself.
Because the disc is extremely optically thick, this energy is temporarily trapped in the centre of the disc
and heats the gas dramatically, sending a weak shock wave out along
the midplane of the disc at a speed of a few km~s$^{-1}$.  However, perpendicular to the disc, the effect is even
more dramatic.  Because there is less material along the rotation axis, the hot gas
finds it easiest to break out in this direction and a bipolar outflow is launched.
Whereas the wave within the disc decays as it travels leaving the
bulk of the disc gravitationally bound, the gas forming the bipolar outflow
has velocities up to 10~km~s$^{-1}$ and travels out into the infalling 
envelope to distances in excess of 50~AU in less than 50 years.  Using two-dimensional
radiation hydrodynamical calculations, \cite{SchTsc2011} have recently reported similar
behaviour, although the outflow in their simulations is less bipolar.  
By disguarding the central regions of one of their calculations, they were able to follow the outflow 
for a few tens of thousands of years and found that after reaching $\approx 500$~AU, the
material fell back in to reform a disc.

\cite{Bate2010} mainly discussed the $\beta=0.005$ calculation as typical example.  Here we
examine how the effect of stellar core formation on the surrounding disc depends on the initial
rotation rate of the molecular cloud core and, thus, on the degree of rotation that the first core/pre-stellar disc
has. Figs.~\ref{images_xyD_OUTFLOW} to \ref{convergence} illustrate the evolution of 
calculations with $\beta=5\times 10^{-4} - 0.01$ following stellar core 
formation.  The $\beta=0$ case is not discussed further
since the situation is as \cite{Larson1969} described it.  This case remains spherically-symmetric, and
although the stellar core irradiates the first core from within, the radiation liberated is not sufficient to
stop the spherically-symmetric accretion flow onto the stellar core.  
We do not discuss the $\beta=0.04$ case following stellar core formation since each of the three
fragments is qualitatively similar to the first core obtained in the $\beta=0.001$ case and, therefore,
we assume that each of these cores would evolve in a similar manner.
 
Figs.~\ref{images_xyD_OUTFLOW} and \ref{images_xzD_OUTFLOW} provide snapshots of the column density in the directions parallel to the rotation axis and perpendicular to the rotation axis, respectively.  The shockwave propagating outwards through each of the first cores/discs is clearly visible in Fig.~\ref{images_xyD_OUTFLOW}.  For the $\beta=5\times 10^{-4}$ and 0.001 cases, the shockwave reaches the outer edge of the disc before the calculations are stopped.  For the $\beta=0.005$ and 0.01 cases, we were not able to follow the calculations this long and the shockwave was only followed $10-15$~AU.  In Fig.~\ref{images_xzD_OUTFLOW}, both the propagation of the shockwaves through the discs and the biploar outflows launched perpendicular to the discs are clearly visible, the latter reaching to distances of 15--35 AU.  The furtherest an outflow was followed was to a distance of 60~AU in the $\beta=0.005$ case, approximately 50 years after the outflow began \citep{Bate2010}.

Each of Figs.~\ref{lines_baro} to \ref{lines_beta0_005} give density, velocity, and temperature profiles, both in the disc plane and perpendicular to the disc plane (i.e. along the rotation axis), at four characteristic times during the
evolution following stellar core formation.  They also provide the radial mass profiles.  Fig.~\ref{lines_baro} shows the evolution of the barotropic $\beta=0.005$ calculation which can be compared with the results from the radiation hydrodynamical calculations in Figs.~\ref{lines_beta0_0005} to \ref{lines_beta0_005} with $\beta=5\times 10^{-4}$ and $0.005$.  We do not provide figures for $\beta=0.001$ and 0.01 since they are qualitatively similar to the cases with $\beta=5\times 10^{-4}$ and 0.005, respectively.

In all calculations, the stellar core forms with a radius of $R_{\rm sc} \approx 0.01$~AU (i.e. $\approx 2 R_\odot$) and a mass of a few M$_{\rm J}$ (e.g. Fig.~\ref{lines_baro}, top-left and top-centre panels).  Using the barotropic equation of state (Fig.~\ref{lines_baro}), the radii of the first core (actually a disc) and stellar core are clearly visible in the top-centre panel which gives the radial velocity profiles just after stellar core formation.  Gas falling on to the first core from the envelope at speeds of $\approx 2-4$~km~s$^{-1}$ is visible at radii of $\approx 10-30$~AU, while gas falling onto the stellar core much more rapidly (up to $7-10$~km~s$^{-1}$) is apparent at radii of $\approx 0.01-0.3$~AU.  In less than a year (second and third rows of Fig.~\ref{lines_baro}), an inner disc builds up around the stellar core (radius $\approx 0.07$~AU in the centre panel of the second row, and $\approx 0.2$~AU in the centre panel of the third row) and the radial extent of the region undergoing dissociation and collapse on to the stellar core and inner disc is reduced.  Several years after the stellar core has formed (bottom row of Fig.~\ref{lines_baro}), the outer edge of the inner disc has completely merged into the outer disc (the remnant of the first core) and the azimuthally-averaged radial velocity in the disc plane is approximately zero out to $\approx 30$~AU.  The formation of an inner disc around the stellar core, which grows in radius as gas with higher angular momentum falls in through the region of molecular hydrogen dissociation, and the eventual merger of the inner and outer discs is just as was first discussed by \cite{Bate1998}, and more recently studied by \cite{MacInuMat2010} and \cite{MacMat2011} who also included magnetic fields.

Using radiation hydrodynamics, the evolution is initially similar to the barotropic cases.  For example, take the $\beta=5\times 10^{-4}$ case (Fig.~\ref{lines_beta0_0005}).  In the top-centre panel, the first core and stellar cores are clearly visible in the radial velocity profiles due to the accretion shocks at their surfaces.  An inner disc builds up around the stellar core as material with higher angular momentum falls inwards and the radial extent of the dissociating, collapsing region decreases (centre panel of the second-row). 

However, as the inner disc grows in radius and begins to merge with remnant of the first core, a bipolar outflow is launched vertically at about 5~km~s$^{-1}$ (dashed line in the centre panel of the third-row of Fig.~\ref{lines_beta0_0005}), while a lower-velocity shock propagates outward along the disc mid-plane at $\approx 1$~km~s$^{-1}$.  All of this evolution occurs within just $4-5$ years of the stellar core forming.  At 20 years after stellar core formation, the outflow has travelled more than 20~AU and the shockwave in the disc has burst out of the edge of the disc (centre panel of the bottom-row of Fig.~\ref{lines_beta0_0005}; see also Fig.~\ref{images_xzD_OUTFLOW}).

The outflow and the shockwave do not vary greatly with the degree of rotation of the first core/pre-stellar disc (c.f. Figs.~\ref{lines_beta0_0005} and \ref{lines_beta0_005}).    In all cases, the outflow is launched (vertically) at a radius of $\approx 0.5$~AU from the stellar core, just before the outer edge of the inner disc merges (in the disc plane) with the remnant of the first core/pre-stellar disc at $\approx 1$~AU and the infalling region of dissociating molecular hydrogen disappears (e.g. the centre panel of the second row in Fig.~\ref{lines_beta0_005}).  The peak speed of the outflow is slightly larger when the first core is rotating more rapidly, ranging from  $\approx 5$~km~s$^{-1}$ for the $\beta=5\times 10^{-4}$ case to $\approx 10$~km~s$^{-1}$ for the $\beta=0.01$ case.  The outflows have been followed for long enough in the $\beta=0.001$ and $\beta=0.005$ calculations (to distances of 30 and 60 AU, respectively) that speeds of the outflows were clearly seen to decay (c.f. the dashed lines in the centre panels of the third and fourth rows of Fig.~\ref{lines_beta0_005}).  This outflowing material will not leave the system; it will eventually stall and fall back onto the protostar and disc \citep{SchTsc2011}.  It is worth noting that because the flux-limited diffusion method used to calculate the radiative transfer in the calculations presented here assumes that the material radiates with maximal efficiency, it may overestimate the cooling and hence underestimate the temperatures in the optically-thin outflowing material.  Thus, the outflow may be even stronger with a more accurate method of radiative transfer.

The shockwave propagating outwards through the disc travels with a speed of $1-2$~km~s$^{-1}$ in all cases.  In the $\beta=5\times 10^{-4}$ and $\beta=0.001$ calculations, the shockwave is followed until after it has reached the edge of the first core/pre-stellar disc.  The combination of the outflow and the outward propagating wave in the disc plane evacuate most of the gas from within a few AU of the protostar, leaving the stellar core surrounded by a broad torus, rather than a disc.  This is apparent in the column density plots in Fig.~\ref{images_xyD_OUTFLOW}, as well as the mass and density profiles in the bottom rows of Figs.~\ref{lines_beta0_0005} to \ref{lines_beta0_005}.  These holes would eventually fill in again as the disc evolves.
In the low-resolution calculations, the mass resolution is poor and the `holes' around the stellar cores are in fact holes (devoid of SPH particles).  However, in the high resolution $\beta=0.001$ and $\beta=0.005$ cases (which each use $3\times 10^6$ particles), it is found that a small amount of gas does remain within the holes, and this gas continues to slowly accrete onto the stellar core.

In Fig.~\ref{convergence}, we plot the mass with density $>10^{-4}$~g~cm$^{-3}$
versus time for the $\beta=0.001$ and $\beta=0.005$ cases with different resolutions.  
It can be seen that the mass of the stellar core at which the
feedback dramatically curtails the accretion 
decreases from $\approx 10-11$ down to 6~M$_{\rm J}$ with increased resolution.
For a few years, the stellar cores grow at a rate of $\sim 10^{-3}$~M$_\odot$~yr$^{-1}$
(even with the highest resolutions).  
With low-resolution ($\le 1\times 10^6$ SPH particles),
the accretion onto the stellar cores ceases entirely after the launching of the
outflows due to the formation of the `holes' in the inner few AU of the discs.  
However, using $3\times 10^6$
SPH particles, it is found that the stellar cores do continue to accrete during this time, but at
much reduced accretion rates of $2\times 10^{-5}$~M$_\odot$~yr$^{-1}$ for the $\beta=0.001$
calculation and $1\times 10^{-5}$~M$_\odot$~yr$^{-1}$ for the $\beta=0.005$ calculation
(measured between 20 and 50 years after stellar core formation; Fig.~\ref{convergence}).

Although the rate of convergence with increasing resolution is relatively slow, strong outflows are
launched in all cases (see Appendix A).  The slow convergence is due to the strong interplay between the
energy released by the formation of the stellar core, and the launching of the outflow which 
decreases accretion and, thus, reduces the energy that is released.  If the resolution is poor,
the stellar core accretes more material before the energy released feeds back and manages to stop the accretion,
whereas with higher resolution, the accretion of a small amount of mass can feed thermal
energy into the infalling material more quickly and, thus, inhibit further accretion.

\section{Discussion}
\label{discussion}

\subsection{The lifetime of the first core/pre-stellar disc phase}

As mentioned in Section \ref{rapidly_rotating}, the introduction of radiative transfer and a 
realistic equation of state increases the lifetimes of the first core phase compared to that obtained
using our barotropic equation of state by factors of 1.5--3 (Fig.~\ref{first_core_time}).  This is important, because the longer the 
first core phase lasts, the more easy it will be to observe.   
The lifetimes obtained using radiative transfer range from $\approx 400$ years (with no rotation)
to $\approx 3000$ years (for $\beta=0.01$), whereas using the barotropic equation of state
they ranged from $\approx 100$ to $\approx 1500$ years.  As discussed above, 
the difference is due to the higher temperatures
(and thus higher pressures) that are obtained using the realistic physics rather than the barotropic
equation of state, which slows the evolution towards the second collapse phase.  This lengthening
of the lifetimes with radiation hydrodynamics compared to barotropic calculations was also 
seen by \cite{Tomidaetal2010a}.  The
lengthening occurs for both the non-rotating first cores, and in the cores that undergo 
non-axisymmetric instabilities.
In the latter, once the first cores or pre-stellar discs develop the bar instability they very
quickly evolve to high central densities and temperatures and undergo collapse to stellar densities
when computed using the barotropic equation of state
(Fig.~\ref{first_core_time}, 4th and 5th panels across where the maximum density increases rapidly 
from $10^{-11}$ to $10^{-7}$~g~cm$^{-3}$, and the maximum temperature from 100 to 2000~K).
Using the more realistic physics, this evolution takes 3--8 times longer.  This is
because the hotter temperatures weaken the strength of the spiral arms generated by the
instability (c.f.\ the third rows of Figs.~\ref{images_D_barotropic} and \ref{images_xyD})
and, thus, decrease the rate of angular momentum transport within the pre-stellar disc.
The removal of the angular momentum from the central regions of the disc removes rotational support 
from the gas \citep{Bate1998}, allowing it to contract and heat up much more rapidly 
than it would due to accretion alone, eventually triggering second collapse.

The pre-stellar disc phase will be more easily observed in more rapidly-rotating 
molecular cloud cores because the lifetime of the phase is longer.  Resolved observations of 
the pre-stellar disc phase (e.g.\ using ALMA) would also be easier and more interesting for the more rapidly-rotating 
cases because the disc will be much larger (up to $\approx 50-100$~AU in radius rather 
than $\approx 5-10$~AU for rotationally-stable first cores) and substructure may be visible 
(i.e.\ the presence of spiral density waves).  Recently, \cite{SaiTom2011} investigated the observability
of pre-stellar discs using ALMA and concluded that ALMA should be able to image objects in the nearest
star-forming regions. For the particular initial 
conditions used in this paper, the lifetimes of as a function of the initial cloud rotation rate
are plotted in Fig.~\ref{firstcore_properties}.  As pointed out recently by \cite{Tomidaetal2010b}, the
lifetime of the first core phase also depends on the mass of the molecular cloud core such
that the lifetimes are longer in lower-mass molecular cloud cores.  For initial core masses
of $\approx 0.1$~M$_\odot$, the accretion onto the first core/pre-stellar disc is not sufficient to drive the
object into the second collapse phase.  Instead, the central regions of the first core/disc evolve
to higher temperatures primarily through radiative cooling.  
As the first core radiates energy, it contracts and heats up, eventually exceeding 2000~K
when the dissociation of molecular hydrogen begins, but this process can take in excess of
$10^4$ years.  Thus, the first core/pre-stellar disc phase is more likely to be observed for molecular cloud
cores that have higher rotation rates and lower masses.

Despite this, observing the pre-stellar disc phase with ALMA will be challenging.  Lifetimes of even $4\times 10^3-10^4$ years are still only 2--5\% of the estimates of Class 0 lifetimes of $2\times 10^5$~yrs \citep{Hatchelletal2007,Evansetal2009}, so that for every 100 Class 0 objects there should only be a few pre-stellar discs.  The first place to look is likely to be the so-called very low luminosity objects (VeLLOs) with luminosities of less than 0.1~L$_\odot$ \citep[e.g.][]{Youngetal2004,Crapsietal2005,Huardetal2006,Bourkeetal2006}, some of which have recently been
proposed as candidate first cores \citep{Chenetal2010,Enochetal2010}.

\subsection{Disc to stellar mass ratios}

As first discovered by \cite{Bate1998}, a rotating first core actually leads to a disc that 
forms {\it before} the star.  These discs can range from $\approx 5$~AU in radius for 
very slowly rotating first cores \citep[c.f.][]{Larson1969} to $\gsim 100$~AU for cores that undergo
rotational instabilities \citep{MacInuMat2010,Bate2010,MacMat2011}.

Because the disc forms first, there is always a phase of star formation when the 
disc to stellar mass ratio is greater than unity.  For example, in the original calculation of
\cite{Bate1998}, the disc had a mass of 0.08~M$_\odot$ when the stellar core formed with an
initial mass of only 1.5~M$_{\rm J}$, giving a disc to stellar mass ratio $\gsim 50$. 
Although some quite massive discs have been discovered around some young stars
\citep[e.g. the 0.8~M$_\odot$ disc around a 3.5~M$_\odot$ star;][]{Hamidouche2010} such cases
are rare \citep[e.g.][]{Boissieretal2011} and the usual assumption is that the 
circumstellar disc mass is always 
likely to be less than the stellar mass.  While this is almost certainly true later in the evolution
of a protostar (e.g.\ Class I or Class II low-mass objects and Herbig Ae/Be stars), in the very earliest phases the disc to star
mass ratio can greatly exceed unity.

\begin{figure}
\centering \vspace{-0.0cm}
    \includegraphics[width=12.5cm]{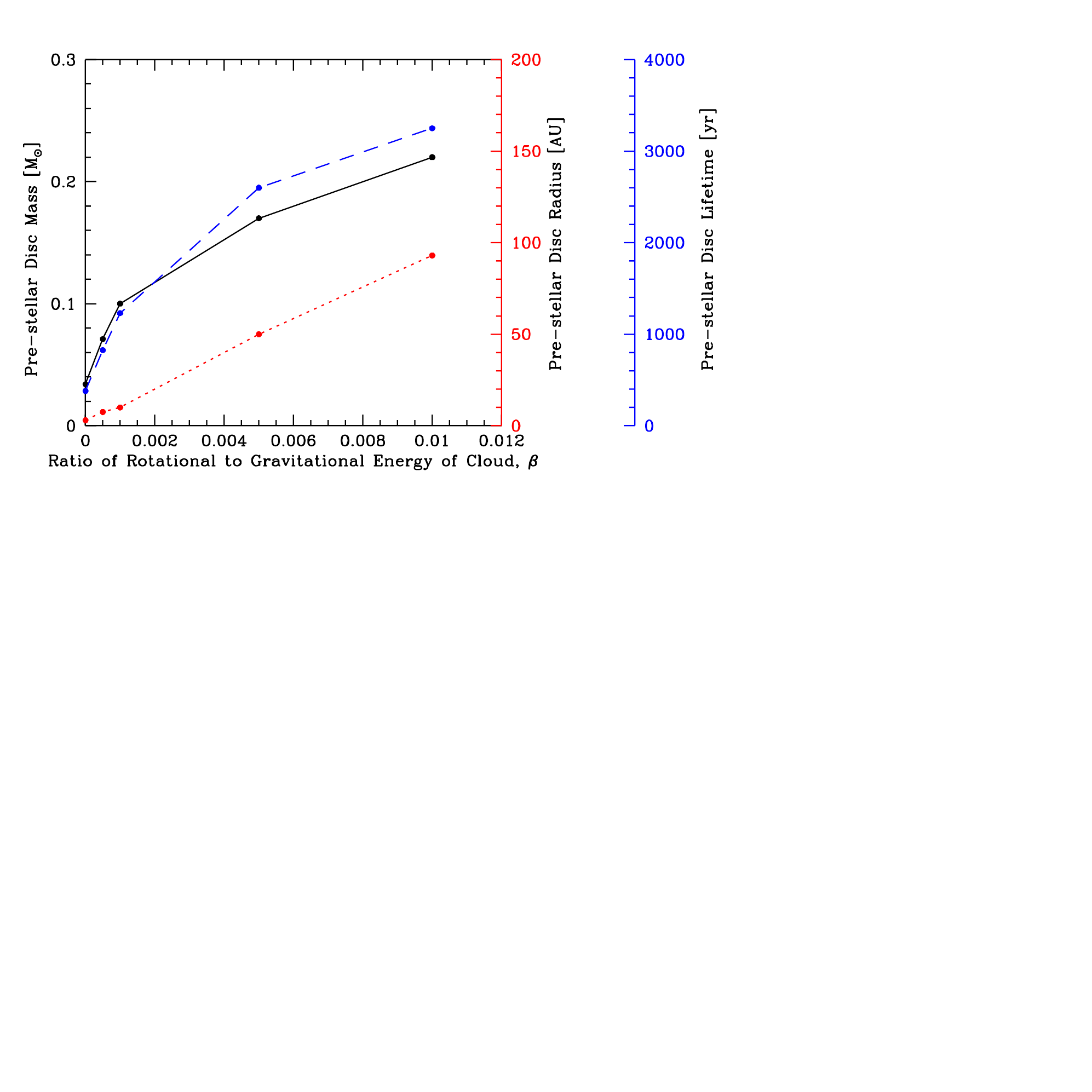}\vspace{-6.8cm}
\caption{The dependence of the pre-stellar disc (first core) properties on the initial rotation rate of the progenitor molecular cloud core, $\beta$.  We plot the masses and radii of the pre-stellar cores, measured just before the second collapse to form the stellar core occurs, and the lifetime of the pre-stellar disc (i.e. the time between the formation of a first hydrostatic core and the formation of the stellar core).  The mass, radius, and lifetime all increase with larger initial molecular cloud core rotation rates.  In the more rapidly-rotating cases, the pre-stellar disc accumulates a significant fraction of the total cloud mass before the stellar core forms.  Note that most rapidly-rotating $\beta=0.04$ case has been excluded because it forms four fragments rather than a single pre-stellar disc.}
\label{firstcore_properties}
\end{figure}

With the initial conditions used in this paper, the disc mass when the stellar core 
forms increases with the rotation rate of the
progenitor molecular cloud core.  This is because a greater fraction of the 1~M$_\odot$ 
molecular cloud core has sufficient angular momentum to settle outside the typical 5~AU
radius of a non-rotating first core.  Because more mass settles away from the centre of the
object, the accretion from the infalling envelope is less effective at increasing the central
temperature (i.e. mass receives support from rotation as well as thermal pressure).  
Thus, the first core phase lasts longer and the masses of the first cores/pre-stellar discs at the time of
stellar core formation range 
from $\approx 0.03$~M$_\odot$ for $\beta=0$ to $\approx 0.22$ M$_\odot$
for $\beta=0.01$ (see Fig.~\ref{firstcore_properties} and the dotted lines in the left panels of 
Figs.~\ref{lines_beta0_0005} to \ref{lines_beta0_005}).  In the most extreme case
we have modelled here, 
when the stellar core forms with an initial mass of $\approx 2$~M$_{\rm J}$, the disc is more than
100 times more massive than the star!  Note that the initial rotation rate of the molecular cloud
core is {\it not} extreme for this case.  \cite{Goodmanetal1993} find that the values of $\beta$
for observed molecular cloud cores range from 0.002 to $>1$, with typical values of $\beta\sim 0.02$.   
Therefore, it is quite possible that when pre-stellar discs
are observed (e.g.\ with ALMA), they could have substantial masses (e.g. up to 1/4 of the mass
of an entire 1-M$_\odot$ molecular cloud core) without a stellar core having formed.

\subsection{Disc fragmentation}

The potential for high disc to stellar mass ratios also implies that some of these discs
may fragment very early in the star formation process.  Indeed, for our $\beta=0.01$ case
(and, of course, the $\beta=0.04$ case),
this is exactly what happens, with the disc fragmenting to produce two additional `first cores'
even before the original first core/disc has undergone second collapse to produce a
stellar core.  The fate of such fragments is of course not clear.  Multiple fragments may
merge with each other before they each undergo second collapse.  They may also spiral in
to the central object and be disrupted before they undergo second collapse.  However,
those that survive will produce the seeds of binary or multiple stellar systems.

As pointed out by \cite{WhiBat2006}, \cite{Krumholz2006}, \cite{KruKleMcK2007}, 
\cite{Bate2009b} and \cite{Offneretal2009}, radiative feedback from newly formed
protostars heats their surrounding gas and inhibits fragmentation.  In star cluster formation
simulations \citep[e.g.][]{Bate2009b,Offneretal2009} this radiative heating dramatically reduces
the number of objects formed compared to using a barotropic equation of state.  Some of this
reduction comes from a reduction in the frequency of disc fragmentation because the discs
are heated and stabilised by the radiation from their central objects \citep{Bate2009b}.

However, as shown here in the $\beta=0.01$ case, fragmentation of the pre-stellar disc resulting
from a rapidly-rotating first core phase can occur {\it before} the central stellar core is formed.
Thus, even if radiative feedback can completely stop disc fragmentation after the stellar core forms,
disc fragmentation {\it before} stellar core formation may be a common route to forming binary
and multiple systems.

\subsection{Implications of radiative feedback following stellar core formation}

The radiative impact of stellar core formation on the surrounding disc, driving a
shock wave through the disc and launching an outflow may have several
wider implications.  These were discussed by \cite{Bate2010}, so we only briefly
list them here.  First, in future studies it will be important
to examine the relative roles of radiative heating due to stellar core formation and
magnetic fields in the launching of a protostellar jet. 
Second, although the outburst and the dramatic decrease in the stellar core
accretion rate discussed in this paper is a transient associated with stellar core formation, 
it might reoccur if the accretion rate onto the stellar core increases again and, thus, may potentially
be a source of episodic accretion and outbursts.  Third, such outbursts may contribute to
the production of the chondrules found in meteorites \citep{Grossmanetal1988}.

\section{Conclusions}
\label{conclusions}

We have presented results from three-dimensional radiation 
hydrodynamical calculations that follow the collapse of a molecular 
cloud core through the formation and evolution of the first hydrostatic
core and beyond the formation of the stellar core.  We find the evolution 
before the formation of the stellar core is qualitatively similar to that found 
in the past using barotropic equations of state.  As found in past studies,
the evolution of the first hydrostatic core depends on the initial rotation
rate of the cloud.  Non-rotating or slowly rotating clouds produce 
small spherical or weakly flattened first cores with radii of $\approx 5-10$~AU.
More rapidly rotating clouds ($\beta \approx 0.001 - 0.01$) produce
highly-oblate rapidly-rotating first cores that undergo dynamic rotational
instabilities to produce large discs (radii $\approx 10-100$~AU) 
before the formation of a stellar core.  These objects are better described 
as pre-stellar discs than first cores.  Yet more rotation typically
results in fragmentation, via a torus shaped first core if the initial cloud
is axisymmetric.  Quantitatively, using radiation hydrodynamics with a realistic equation
of state produces first cores that are somewhat hotter, slightly more stable to
rotational instabilities, and longer-lived (by factors of $1.5-3$) than
using our barotropic equation of state.

The masses and radii of the pre-stellar discs produced before stellar core
increase with the initial rotation rate of the molecular cloud core.
Their lifetimes also increase with rotation rate, and for 
lower-mass molecular cloud cores \citep[see also][]{Tomidaetal2010b}.
With high rotation rates, these pre-stellar discs can have radii 
up to $\approx 100$~AU and contain
in excess of 0.2~M$_\odot$ of gas (up to $1/4$ of the mass of the entire
molecular cloud core) and exist for several thousands of years
before stellar core formation.  Such objects may be resolvable
by ALMA.  Since initial mass of the stellar core which
forms within these discs is just a few Jupiter-masses, the disc-to-star
mass ratios can exceed 100 for a short period of time. 

Fragmentation of a pre-stellar disc prior to stellar core formation
may be an effective way to produce binary and multiple star systems
because the radiative feedback from the stellar core which may inhibit
later fragmentation of a massive disc is absence.

After the formation of a stellar core within the disc,
dissociating gas from the inner regions of the disc falls
towards the stellar core and builds a smaller disc around the
stellar core.  This inner disc increases with radius as material
with higher angular momentum falls in, and eventually merges
with the inner edge of the outer disc.  

Although this inner/outer disc structure and the earlier evolution of the 
first core phase is similar for barotropic and radiation hydrodynamical calculations, 
we find that the subsequent evolution is qualitatively different in radiation 
hydrodynamical calculations.
In barotropic calculations, the formation of the stellar core deep
inside the first core (or pre-stellar disc) has no effect on the surrounding disc because
the temperature of the gas is simply set by the density of the gas.
However, with radiation hydrodynamics, the energy released by the formation
of the stellar core within the optically thick disc is similar to the binding 
energy of the disc.  This heats region surrounding the stellar core and,
at approximately the same time as the inner disc and outer discs merge, 
a shock wave forms in the merger region and propagates outwards through the disc, 
while a bipolar outflow is launched perpendicular to the rotation axis,
dramatically decreasing the accretion rate on to the stellar core.
The properties of the outflow do not seem to vary greatly with the
initial rotation rate of the molecular cloud core or the pre-stellar disc,
but the speed of the outflow is somewhat greater for higher rotation rates
and the shockwave which propagates outward through the
disc reaches the outer edge of the disc more quickly for smaller discs.
It remains to be seen how the inclusion of magnetic fields alters this
thermally-driven outburst.

\section*{Acknowledgments}

MRB is grateful for the hospitality and support he received from the Monash Centre for Astrophysics while on study leave and where this paper was written up.
The calculations for this paper were performed on the University of Exeter Supercomputer, 
a DiRAC Facility jointly funded by STFC, the Large Facilities Capital Fund of BIS, and the University of Exeter.
Some figures were produced using the publicly available SPLASH visualisation software \citep{Price2007}.
This work, conducted as part of the award ``The formation of stars and planets: Radiation hydrodynamical and magnetohydrodynamical simulations" made under the European Heads of Research Councils and European Science Foundation EURYI (European Young Investigator) Awards scheme, was supported by funds from the Participating Organisations of EURYI and the EC Sixth Framework Programme.

\bibliography{mbate}

\begin{figure}
\centering \vspace{-0.5cm}
    \hspace*{-0.15cm}\includegraphics[width=12.5cm]{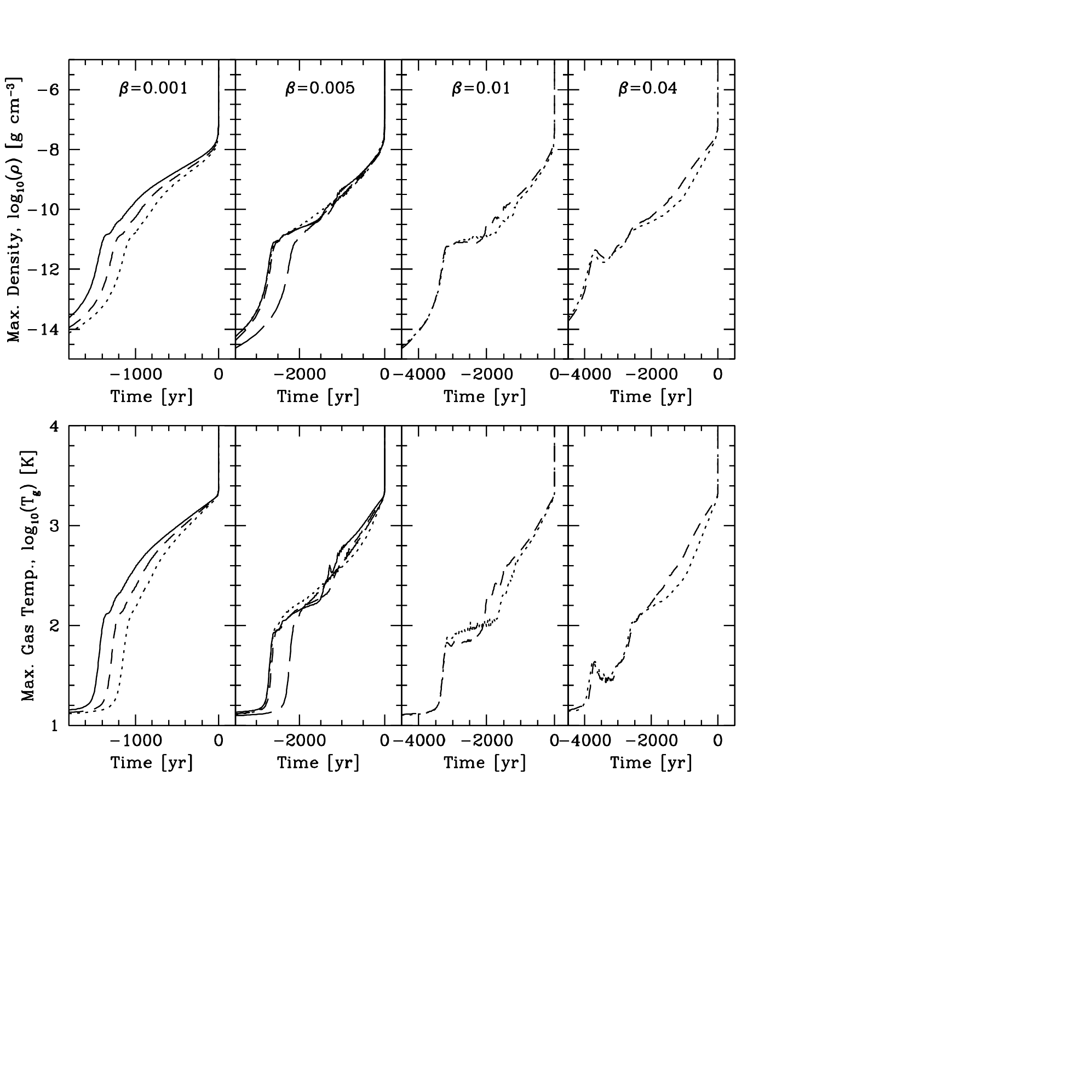}\vspace{-3.5cm}
\caption{Examining the numerical convergence of the lifetime and evolution of the first core/disc.  We plot the time evolution of the maximum density (upper panels) and gas temperature (lower panels) during the radiation hydrodynamical calculations of the collapse of molecular cloud cores with different initial rotation rates.  From left to right, the different panels are for cloud cores with $\beta=0.001,0.005,0.01,0.04$. The different line types are for calculations performed using $1\times 10^5$ (long-dashed line), $3\times 10^5$ (dotted lines), $1\times 10^6$ (short-dashed lines), and $3\times 10^6$ (solid lines) SPH particles.   Time is set to zero at the end of the second dynamic collapse phase when the density reaches $10^{-3}~{\rm g}~{\rm cm}^{-3}$ which allows the length of the first hydrostatic core phases to be compared.  It can be seen that the lifetime of the first core phase is well converged for all except the lowest resolution calculation for the $\beta=0.005$ case, and the $\beta=0.001$ case.  This is because the lifetime of the first core/disc is determined by the dynamic rotational instability for $\beta\ge 0.005$, whereas for $\beta=0.001$ viscous evolution probably plays a significant role.}
\label{first_core_time_convergence}
\end{figure}

\begin{figure*}
\vspace{-0.5cm}
	\hspace*{3.68cm}\includegraphics[width=18.0cm]{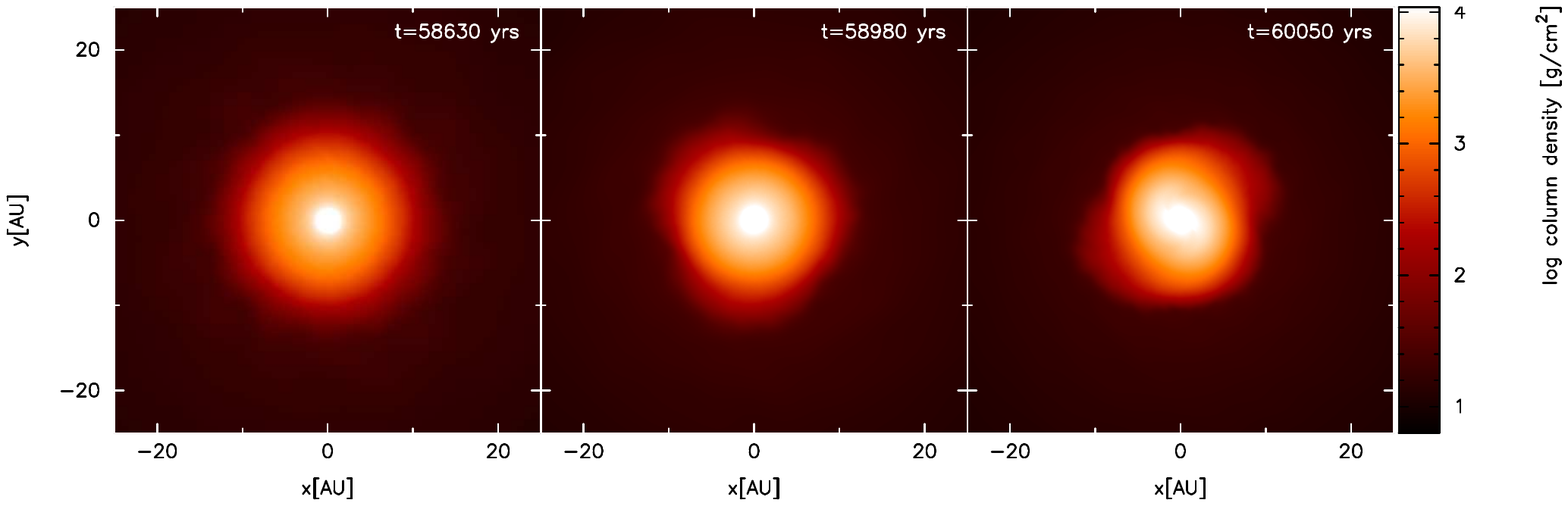}\vspace{-9.9cm}
    \includegraphics[width=18.0cm]{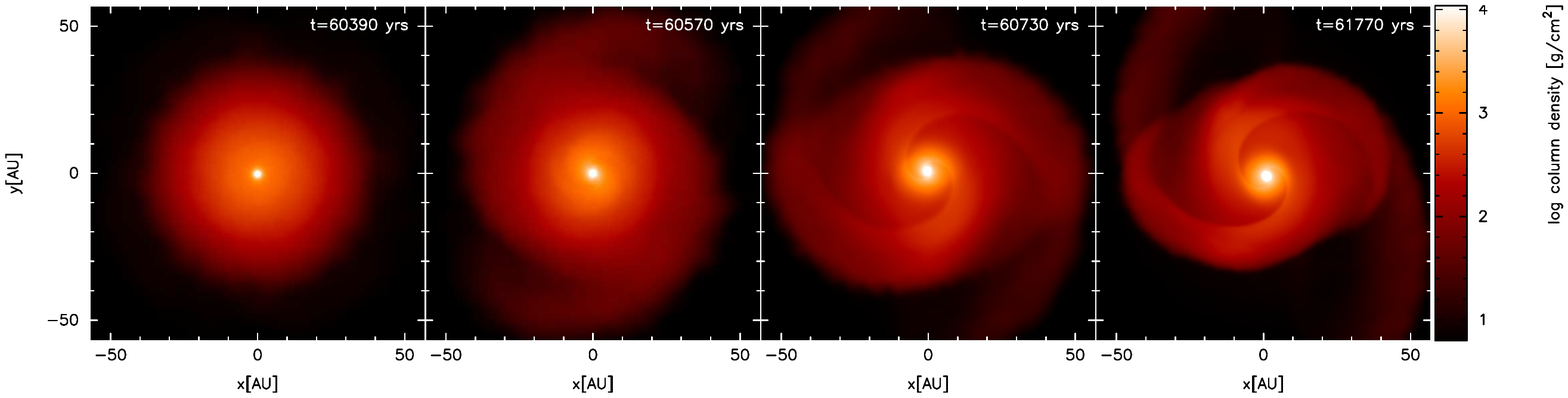}\vspace{-9.9cm}
    \hspace*{3.68cm}\includegraphics[width=18.0cm]{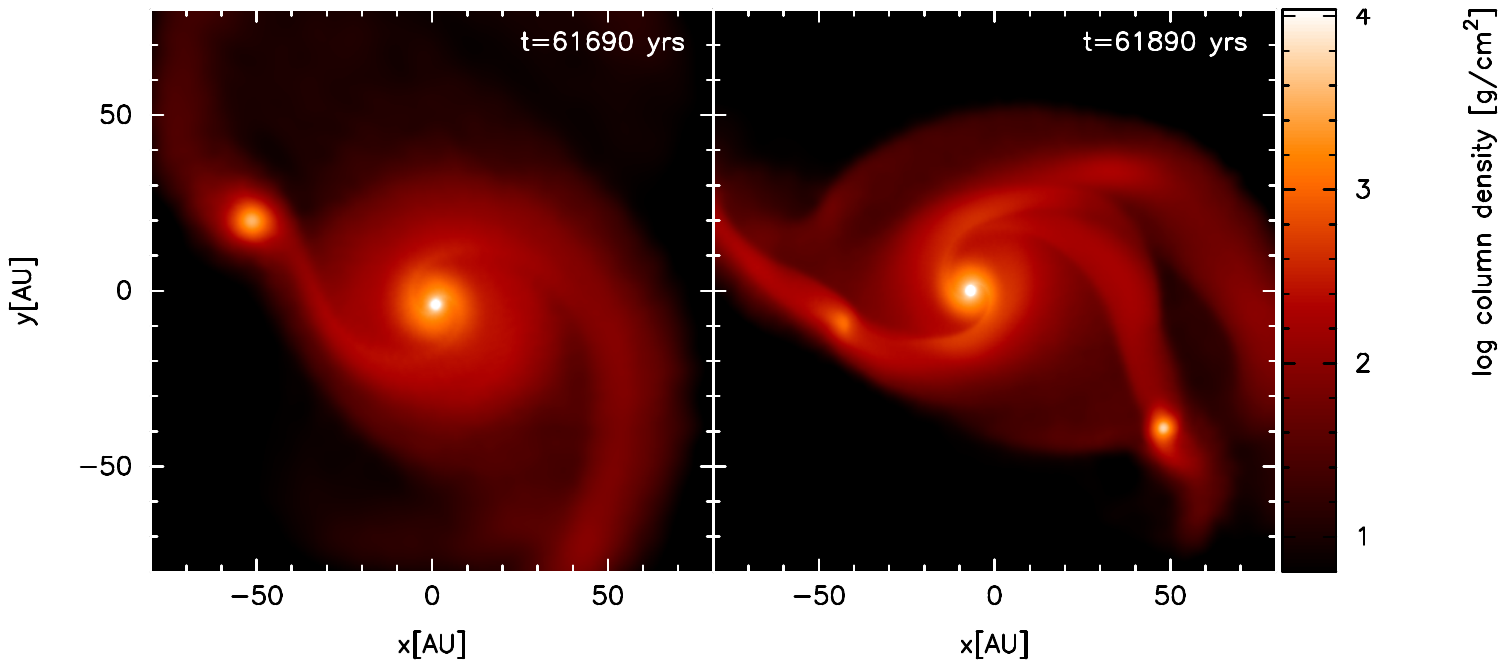}\vspace{-9.9cm}
    \hspace*{3.68cm}\includegraphics[width=18.0cm]{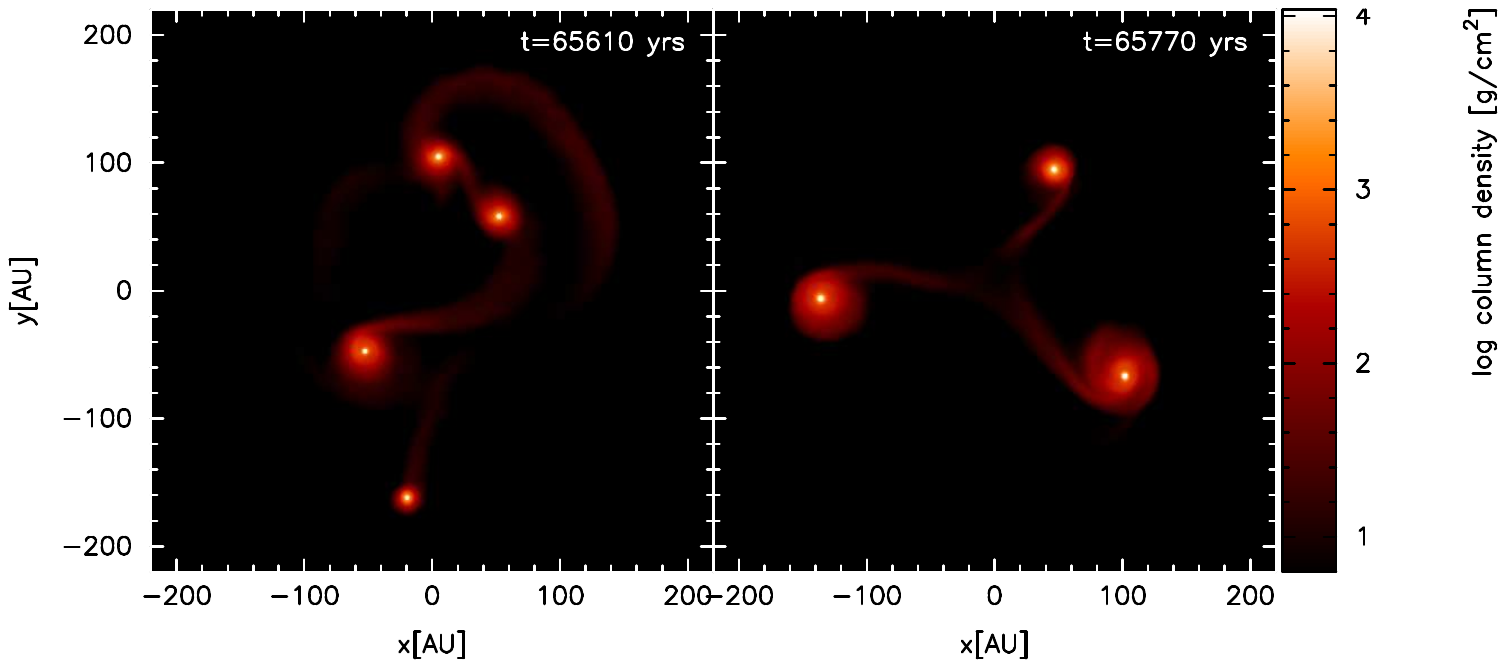}\vspace{-9.0cm}
\caption{Snapshots of the column density viewed along the rotation axis at the point of stellar core formation (defined as when the maximum density reaches $1\times 10^{-3}$~g~cm$^{-3}$.  From top to bottom, the different rows are for molecular cloud cores with $\beta=0.001,0.005,0.01,0.04$. Note that the spatial scale is different for each row.  From left-to-right, the columns are for calculations performed with $1\times 10^5$, $3\times 10^5$, $1\times 10^6$, and $3\times 10^6$ SPH particles.  For each rotation rate, the evolution is qualitatively the same when using $3\times 10^5$ particles or more.
}
\label{images_D_convergence}
\end{figure*}

\begin{figure*}
\vspace{-0.5cm}
	\includegraphics[width=18.0cm]{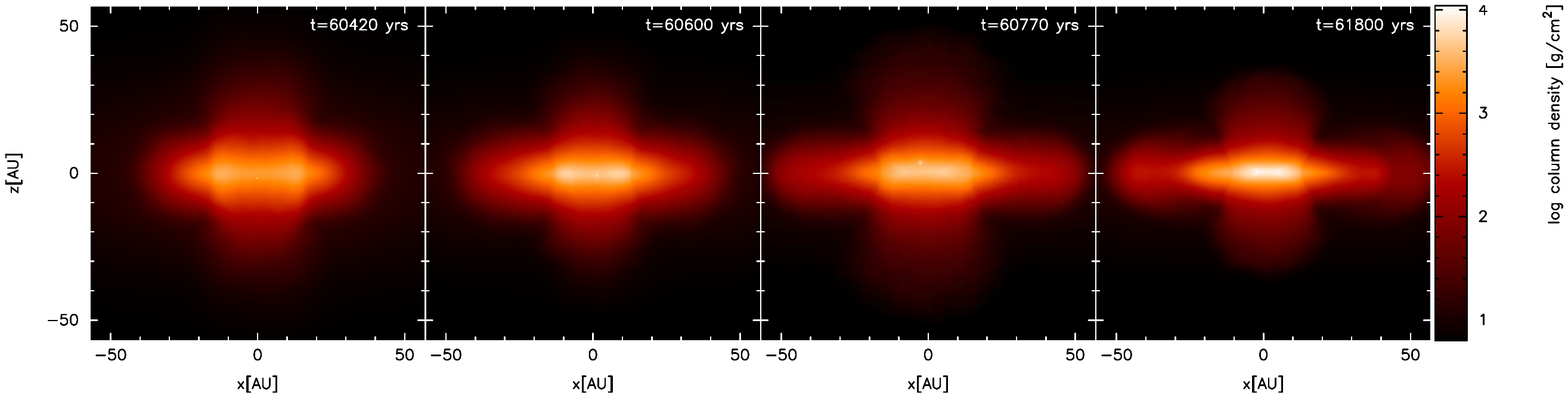}
\vspace{-9.0cm}
\caption{Snapshots of the column density viewed perpendicular to the rotation axis 25 years after stellar core formation  (defined as when the maximum density reaches $1\times 10^{-3}$~g~cm$^{-3}$) in a $\beta=0.005$ molecular cloud core.  From left to right, the different panels are for radiation hydrodynamical calculations performed with $1\times 10^5$, $3\times 10^5$, $1\times 10^6$, and $3\times 10^6$ SPH particles.  The bipolar outflow and the shockwave propagating through the disc are similar in all cases, regardless of the resolution.}
\label{outflow_convergence}
\end{figure*}

\section*{Appendix A: Numerical convergence of the radiation hydrodynamical calculations}
\label{appendix}

As mentioned in Section \ref{initialconditions}, for the different initial conditions and equations of state discussed in this paper, most of the calculations were performed with several different numerical resolutions to check for numerical convergence of the results.  The numbers of SPH particles used were $1\times 10^5$, $3\times 10^5$, $1\times 10^6$, and $3\times 10^6$ particles, although as summarised in Table \ref{resolutions} not all cases were performed with all resolutions.  The only type of calculation to be performed with all four resolutions was the $\beta=0.005$ case with radiation hydrodynamics, but the evolution of three other initial conditions performed with radiation hydrodynamics were studied with two ($\beta=0.01$ and 0.04) or three ($\beta=0.001$) different resolutions.  In this appendix, we discuss how the results depend on numerical resolution.

As numerical resolution was increased, all of the clouds took slightly longer to collapse.  However, more important for the results presented in the main sections of this paper is how the lifetime and evolution of the first core or pre-stellar disc phase depends on numerical resolution.  In Figure \ref{first_core_time_convergence}, we plot the evolution of maximum density and temperature versus the time before stellar core formation (defined as when the maximum density reaches $10^{-3}$~g~cm$^{-3}$) for the radiation hydrodynamical calculations with $\beta=0.001-0.04$.  The lifetimes of the pre-stellar disc phase are all well converged for $\ge 3\times 10^5$ particles, with the exception of the $\beta=0.001$ case.  For the $\beta=0.005-0.04$ cases, the lifetime of the first core/pre-stellar disc phase depends on the development of dynamical non-axisymmetric instabilities due to the high ratio of the rotational to gravitational potential energies.  In particular, the transport of angular momentum via gravitational torques from the spiral density waves removes rotational support from the gas and drives the object to the second collapse phase much more quickly that it would evolve due to accretion and radiative cooling alone (see Section \ref{rapidly_rotating}).  However, the $\beta=0.001$ case is rotationally stable.  Therefore, it evolves to the second collapse phase through accretion and radiative cooling, but also through angular momentum transport by numerical shear viscosity.  Numerical shear viscosity will transport angular momentum outwards, reducing the rotational support of the gas in the centre of the first core, driving it more quickly to higher densities and temperatures and, thus, hastening the onset of the second collapse.  As the resolution is increased, the numerical shear viscosity decreases.  Therefore, it is most likely that the longer lifetimes of the pre-stellar disc phase for the $\beta=0.001$ case with increasing resolution is due to the effect of the numerical shear viscosity in SPH.  We also note that the dominant term in the numerical viscosity (the $\alpha_{\rm v}$ term) is first-order, which explains why the convergence with increasing resolution is quite slow (left panels of Figure \ref{first_core_time_convergence}).  Thus, for the first cores which do not undergo dynamical rotational instabilities, the lifetimes obtained in this paper should be treated as lower limits.

In Fig.~\ref{images_D_convergence}, we plot images of the first cores/pre-stellar discs at the time of stellar core formation for the $\beta=0.001-0.04$ radiation hydrodynamical cases obtained using the different numerical resolutions.  For the $\beta=0.001$ case, the first core is right at the boundary of rotational stability, with a very weak spiral instability visible in the highest resolution case.  For the $\beta=0.005$ case, all resolutions except the lowest ($1\times 10^5$ particles) resolve the rotational instability.  As the resolution is increased, the sharpness of the spiral density waves improves and the pre-stellar disc becomes slightly smaller, but the structure definitely seems to be converging.  For the $\beta=0.01$ case, there are only two resolutions.  Both produce a central object surrounded by a disc of radius $\approx 90$~AU which fragments.  The higher resolution case forms two fragments, while the lower resolution case only has one fragment.  Thus, while the number of fragments has not converged, the size of the pre-stellar disc and the fact that it does fragment seem to be robust.  Finally, both resolutions of the $\beta=0.04$ case produce ring-type first cores which fragment into four separate cores.  However, with four objects the subsequent evolution becomes chaotic.  In the $3\times 10^5$ particle calculation, the four fragments survive (at least until the first undergoes second collapse), but in the $1\times 10^6$ particle calculation two of the objects merge producing a triple system (see the bottom rows of Fig.~\ref{images_D_convergence} and \ref{images_xyD}).

Following stellar core formation, as discussed in detail in Section \ref{stellar_core_formation}, a bipolar outflow is launched perpendicular to the disc and a shockwave propagates outwards along the plane of the disc.  These outflows and shockwaves occur in all of the radiation hydrodynamical simulations, with the exception of the spherically-symmetric $\beta=0$ case.  This result does not appear to depend on resolution (see Figure \ref{outflow_convergence}).  However, as discussed in Section \ref{stellar_core_formation} and illustrated in Fig.~\ref{convergence}, the initial mass of the stellar core and its accretion rate are affected by resolution.

In summary, we conclude that the lifetimes of the first cores/pre-stellar discs are well converged for initial conditions that produce rotationally-unstable pre-stellar discs, but that for more slowly-rotating cores (i.e. $\beta=0.0005$ and $\beta=0.001$) the lifetimes are lower-limits.  We also find that the qualitative outcomes (rotationally-stable first core, pre-stellar disc with spiral arms, fragmenting pre-stellar disc, and torus) are numerically-converged but the details (e.g.\ how many fragments) may not be.

\end{document}